\documentclass[12pt]{article}
\usepackage{fullpage}
\usepackage{amsmath}
\usepackage{amssymb}
\usepackage{amsthm}
\usepackage{dsfont}
\usepackage{latexsym,amsfonts,epsfig, psfrag}
\usepackage{enumerate}

\usepackage[mathscr]{euscript} 

\def\be{\begin{equation}}
\def\ee{\end{equation}}
\def\ba{\begin{eqnarray}}
\def\ea{\end{eqnarray}}

\def\nn{\nonumber}

\begin{document}

{\renewcommand{\thefootnote}{\fnsymbol{footnote}}
\medskip

\begin{center}
  {\LARGE Semiclassical States on Lie Algebras}\\
 Artur Tsobanjan\footnote{e-mail address: {\tt artur.tsobanjan@gmail.com}}
  \\
  \vspace{0.5em}
 King's College, 133 North River Street, Kingston, PA 18702, USA\\ and\\
Institute for Gravitation and the Cosmos,
  The Pennsylvania State
  University,\\
  104 Davey Lab, University Park, PA 16802, USA\\
  \vspace{1.5em}
\end{center}
}
\begin{abstract}
The effective technique for analyzing representation-independent features of quantum systems based on the semiclassical approximation (developed elsewhere), has been successfully used in the context of the canonical (Weyl) algebra of the basic quantum observables. Here we perform the important step of extending this effective technique to the quantization of a more general class of finite-dimensional Lie algebras. The case of a Lie algebra with a single central element (the Casimir element) is treated in detail by considering semiclassical states on the corresponding universal enveloping algebra. Restriction to an irreducible representation is performed by ``effectively'' fixing the Casimir condition, following the methods previously used for constrained quantum systems. We explicitly determine the conditions under which this restriction can be consistently performed alongside the semiclassical truncation.
\end{abstract}

\section{Introduction}

Effective equations for quantum systems developed in~\cite{EffAc} and~\cite{EffEq2} allow quick access to the behavior of a quantum system in the semiclassical regime without performing a formal Hilbert space construction. The usefulness of this approach has been showcased for non-harmonic quantum systems~\cite{Boj_Brahm_Nels} and cosmological models~\cite{EffCosm}. However, this method was originally developed for the situations where the fundamental kinematical variables of the system to be analyzed form the so-called canonical algebra (sometimes called Weyl-algebra). In order to make it applicable to a more general class of cosmological models or, more ambitiously, to gauge theories, it needs to be extended to the situations, where the defining variables form a more general Lie algebra. The major new feature of such observable algebras is the presence of ``redundancy'' conditions expressed by Casimir operators. In~\cite{Bojowald_sl2c_eff} these redundancy conditions are imposed following the methods of ``effective quantum constraints'' developed in~\cite{EffCons} and~\cite{EffConsRel}. While, by itself such a procedure is sensible, it remained to be demonstrated that it can be consistently implemented alongside the semiclassical truncation of the system.

Establishing the consistency of combining semiclassical truncation and removing redundancy is the main result of this report. We work in the setting of a general finite-dimensional Lie algebra and perform explicit computations for the case of a single non-trivial center-generating element. The discussion is organized into three sections. Section~\ref{sec:QLieAlg} describes our approach to states on a Lie algebra and sets up both the semiclassical truncation and the reduction by the central element, summarizing the detailed results from subsequent sections. Section~\ref{sec:PBOrder} establishes the consistency of the reduction in the degrees of freedom and the quantum Poisson structure that they inherit from the Lie algebra during the semiclassical truncation. Section~\ref{sec:Constraints} uses explicit counting to determine the conditions under which the  correct number of redundancy conditions remain after the semiclassical truncation.

\section{Semiclassical phase-space of a Lie algebra} \label{sec:QLieAlg}

We will focus our attention on the case of an associative unital algebra $\mathcal{A}_{kin}$, which is the universal enveloping algebra of some Lie algebra $\mathfrak{a}$, such that the center of $\mathcal{A}_{kin}$ is generated by a single independent non--trivial element $\hat{P}$, for example, the Casimir element in the case of semisimple Lie algebras. The case where the center of  $\mathcal{A}_{kin}$\ has several independent generators can be treated in a similar way, by considering one generator at a time. Since our technique has been developed to complement the usual representation-based methods of quantum mechanics, the space of all linear states on $\mathcal{A}_{kin}$\ is of particular interest. Although $\mathcal{A}_{kin}$\ is infinite-dimensional as a linear space, to a given semiclassical order its space of states can be captured by a finite set of (non-linear) functions with the original Lie algebra structure giving rise to a (typically degenerate) Poisson bracket. The degree of freedom redundancy associated with $\hat{P}$\ can be accounted for  alongside this truncation in the form of polynomial conditions on the non-linear functions mentioned above.

In the rest of this section we first give some general quantum-mechanical motivation for considering states on the quotient space $\mathcal{A}_{kin}/ \mathcal{A}_{kin} \hat{C}$, where $\hat{C}=\hat{P}-r\hat{\mathbf{1}}$; we refer to passing to this quotient as ``reduction''. We then briefly overview the quantum phase-space generated by linear states on $\mathcal{A}_{kin}$. In~\ref{sec:Semiclassical} we describe the semiclassical truncation on this quantum phase-space. We conclude this section by considering the conditions under which the semiclassical truncation and reduction in the degrees of freedom give consistent results. The details of our argument for consistency are presented in the sections that follow.

\subsection{Quantum system as an algebra}
Quantum description of a finite--dimensional mechanical system can be thought of as a particular irreducible representation of a chosen (finite--dimensional) Lie algebra of phase-space functions that describe freedoms of the system classically. For example, the classical degrees of freedom of a particle moving in one dimension can be captured by its position $x$ and momentum $p$, subject to the canonical Poisson relation $\{x, p\}=1$. Within the usual ``Schr\"odinger'' quantum mechanics of a particle in one dimension the degrees of freedom are captured by the canonical algebra defined by the commutation relation $[\hat{x}, \hat{p}] = i\hbar \hat{\mathbf{1}}$, where $\hat{x}$\ and $\hat{p}$\ are now differential operators, generating an infinite-dimensional associative algebra. The slightly more general situation of a Lie algebra with a central element in its corresponding universal enveloping algebra arises when quantizing a mechanical system with a phase-space $\Gamma$\ that possesses an over-complete set of
coordinates $\{x_i\}$, $i=1, 2, \ldots M$, such that:
\begin{enumerate}[(i)]
\item  $\{x_i\}$\ resolve the points of $\Gamma$;

\item  they form a Lie algebra under the Poisson bracket: ${\displaystyle \{x_i, x_j\} = \alpha_{ij}^{\ \ k} x_k}$, where $\alpha_{ij}^{\ \ k}$\ are structure constants (typically real) and summation over repeated indices is implied;

\item the over-completeness is expressed by a single polynomial
relationship $P(x_1, \ldots, x_M) = const.$, with $\{P, x_i\} = 0$ for all coordinate functions $x_i$.
\end{enumerate}
A simple example of such a system is the sphere $\Gamma
\cong S^2$, with the area form providing the symplectic structure. 
Viewed as embedded in the Euclidean space $\mathbb{R}^3$\ it has an
over-complete set of Cartesian coordinates $x$, $y$, $z$, that obey
the $su(2)$\ Lie algebra (of rotations about the three coordinate axes) with respect to the corresponding Poisson bracket. The
redundancy is expressed by the Casimir element of $su(2)$, we
have $x^2+y^2+z^2 = const.$\ Another example is provided by the
bouncing cosmological model introduced in~\cite{Bojowald_sl2c_1}~and~\cite{Bojowald_sl2c_2}, which
is based on the Lie algebra $sl(2, \mathbb{C})$. More generally, our hope is to extend the effective method to the quantum treatment of gauge fields, whose degrees of freedom at each spatial point are captured by a finite--dimensional Lie algebra. 

By analogy with the standard notation for operators in quantum mechanics, we denote the generators of the universal enveloping algebra by $\hat{x}_i$, using ``\ $\widehat{\ }$ '' to distinguish elements of $\mathcal{A}_{kin}$\ emphasizing that multiplication is no longer abelian as it was in classical mechanics. The universal enveloping algebra is the collection of all (typically complex) polynomials in $\hat{x}_i$\ modulo the canonical commutation relations (CCRs)\footnote{The factor of $\hbar$\ appearing in the CCR can be treated as a formal parameter keeping track of the number of commutators taken. It does not change the Lie algebra structure and can be easily absorbed by using $\hat{\chi}_i = \hat{x}_i /\hbar$\ as generators.}
\begin{equation} \label{eq:CCR}
[\hat{x}_i, \hat{x}_j] := \hat{x}_i\hat{x}_j - \hat{x}_j\hat{x}_i = i\hbar \alpha_{ij}^{\ \ k} \hat{x}_k \ .
\end{equation}
This is one way of enforcing the so-called ``Dirac's condition'' for quantization, that establishes the correspondence between the quantum commutator and the classical Poisson bracket. Here $\hbar$\ introduces the quantum scale---we will later explicitly treat it as a small quantity for the purpose of semiclassical truncation.

The universal enveloping algebra is Lie-homomorphic to
any Hilbert space representation of such a system and therefore contains its representation-independent features. We will refer to $\mathcal{A}_{kin}$\ as the \emph{kinematical} algebra of the system because it describes the system before the redundacy conditions and its dynamical properties (e.g. selecting a prefered Hamiltonian element) are imposed. Here we will not worry about the dynamics, concentrating instead on the semiclassical method of imposing the redundancy conditions. Since any physical properties of interest must preserve the true phase-space defined by $P(x_1, \ldots, x_M) = const.$, we expect that the true quantum degrees of freedom are given by the quotient $\mathcal{A}_{kin}/\mathcal{A}_{kin} \hat{C}$, where $\hat{C} = \hat{P} - r\hat{\mathbf{1}}$, $\hat{P}$\ is the quantization of $P$\ and the generator of the center of $\mathcal{A}_{kin}$\ and $r$\ is some real-valued constant. In a semisimple Lie algebra, $\hat{C}$\ would fix the value of the Casimir charge. In reference to this class of systems we will refer to it as the \emph{Casimir constraint}\footnote{In the case of a compact phase-space defined by the condition $P(x_1, \ldots, x_M) = const.$, $r$\ in general cannot take arbitrary real values, but must be set to one of the allowed quantum numbers for consistency. However, in the semiclassical regime the phase space volume, and hence also $r$, must be sufficiently large compared to the gaps between their possible discrete values to allow existence of ``sharply peaked'' states. Thus in what follows the precise value $r$\ will not be important.}.
For example, in the above example of a spherical phase-space, the Casimir constraint fixes the size of the phase space $\hat{C} = \hat{x}^2 + \hat{y}^2 + \hat{z}^2 - R^2\hat{\mathbf{1}}$. Since $\hat{C}$\ commutes with every element of
$\mathcal{A}_{kin}$, the quotient is in fact a well-defined associative unital algebra. In the following subsections we briefly describe the setup for creating a semiclassically truncated version of this quotient algebra.

\subsection{The quantum phase-space} \label{sec:QPhaseSpace}

For our purposes, states on  $\mathcal{A}_{kin}$\ are linear maps from  $\mathcal{A}_{kin}$\ to complex numbers. The space of all such maps that, in addition, assign 1 to the identity element $\hat{\mathbf{1}}$, will be referred to as the quantum phase-space $\Gamma_Q$\ of the system. \footnote{In algebraic constructions of representations one typically imposes the positivity condition on states, however in the case of $\Gamma_Q$\ we make no such restriction to begin with. In particular applications of the effective technique positivity conditions are often imposed once constraints on $\mathcal{A}_{kin}$ are accounted for. In the language of constrained quantum mechanics, the states on the full algebra $\mathcal{A}_{kin}$\ can be thought of as kinematical.} By the natural duality, each element of  $\mathcal{A}_{kin}$\ is also a linear function on $\Gamma_Q$. We will denote the function induced by $\hat{a} \in  \mathcal{A}_{kin}$\ as $\langle \hat{a} \rangle$\ by direct analogy with ordinary quantum mechanics, where the expectation value of an operator is a specific example of such a function. Indeed, we will explicitly refer to functions of the type $\langle \hat{a} \rangle$\ as \emph{expectation values}. Evidently, if $\varphi$\ is an element of $\Gamma_Q$, then there is a complex number $\varphi(\hat{a})$\ that defines the value taken by the expectation value function $\langle \hat{a} \rangle$\ at the quantum phase-space point $\varphi$. It is straightforward to establish that this identification between the elements of $\mathcal{A}_{kin}$\ and expectation value functions on $\Gamma_Q$\ is linear, e.g. $\langle \hat{a} + \hat{b} \rangle = \langle \hat{a} \rangle + \langle \hat{b} \rangle$. Additionally, the expectation value is ``normalized\rq{}\rq{}, that is $\langle \hat{\boldsymbol{1}}\rangle = 1 \ $. 

Importantly, a subset of expectation value functions resolves the points of $\Gamma_Q$. This follows as $\mathcal{A}_{kin}$\ has a (countable) linear basis, for example polynomials of the form $\hat{x}_1^{i_1}\hat{x}_2^{i_2} \ldots \hat{x}_M^{i_M}$, where $M$\ is the dimension of the Lie algebra. Therefore, an element of $\Gamma_Q$\ is completely defined by the values it assigns to such a basis, meaning that the expectation value functions induced by any such basis form a complete set of coordinates on $\Gamma_Q$.

Since $\Gamma_Q$\ is the (vector space) dual of an algebra, it possesses additional structure. Most importantly, the Lie-algebraic structure inherited by $\mathcal{A}_{kin}$\ from the original Lie algebra $\mathfrak{a}$ is passed along to $\Gamma_Q$\ in the form of the quantum Poisson bracket $\{.,.\}_Q$. For a pair of expectation value functions it is defined directly by
\begin{equation}\label{eq:QPoisson}
\left\{ \langle \hat{a}\rangle ,\langle \hat{b} \rangle \right\}_Q := \frac{1}{i\hbar} \left\langle [\hat{a}, \hat{b} ] \right\rangle =  \frac{1}{i\hbar} \left\langle \hat{a} \hat{b} - \hat{b} \hat{a} \right\rangle \ .
\end{equation}
Jacobi identity for the above bracket follows quickly from properties of the commutator in an associative algebra. The bracket extends to non-linear functions on $\Gamma_Q$\ by linearity and Leibniz rule. 

The quantum Poisson bracket allows one to formulate dynamical relations of quantum mechanics in the form closely analogous to the classical Hamiltonian mechanics, which is the original motivation for developing this approach. In particular, the unitary time-evolution of a quantum wave-function, leads to the standard result of quantum mechanics for the corresponding time-evolution of the expectation value of any (time-independent) operator
\[
\frac{d}{dt} \langle \hat{a} \rangle =  \frac{1}{i\hbar} \left\langle [\hat{a}, \hat{H} ] \right\rangle \ ,
\]
where $\hat{H}$\ is the Hamiltonian of the system. In terms of the quantum Poisson bracket, this is simply
\begin{equation} \label{eq:Qevolution}
\frac{d}{dt} \langle \hat{a} \rangle = \left\{ \langle \hat{a}\rangle ,\langle \hat{H} \rangle \right\}_Q
\end{equation}
By linearity and Leibniz rule, this result extends to any function $f$ on $\Gamma_Q$, so that $df/dt = \{f, H_Q \}_Q$, where $H_Q = \langle \hat{H} \rangle$\ is refered to as the \emph{quantum Hamiltonian} in the literature.

Since the degrees of freedom of $\mathcal{A}_{kin}$\ contain redundancy, so does the quantum phase-space $\Gamma_Q$. Algebraic redundancy is (formally) removed by passing to the quotient $\mathcal{A}_{kin}/\mathcal{A}_{kin} \hat{C}$. The true quantum phase-space should therefore be the space of linear states on the above quotient algebra. This can be implemented by passing to the subspace of $\Gamma_Q$\ that vanishes on $\mathcal{A}_{kin} \hat{C}$. In particular, ``physical'' states must satisfy
\[
\langle \hat{a} \hat{C} \rangle = 0 \quad , \quad \forall \hat{a} \in \mathcal{A}_{kin} \ .
\]
This can be systematically implemented through a (countably) infinite set of conditions using any basis on $\mathcal{A}_{kin}$, for example
\begin{equation}\label{eq:CPoly}
\left\langle \hat{x}_1^{i_1}\hat{x}_2^{i_2} \ldots \hat{x}_M^{i_M} \hat{C} \right\rangle = 0 \quad , \quad \forall \ \ i_1, i_2, \ldots i_M \ .
\end{equation}
Consider $\Sigma$, the linear subspace of the phase space $\Gamma_Q$, where all of the above conditions are satisfied. We quickly infer that it is a Poisson submanifold of $\Gamma_Q$, since
\[
\left\{ \langle \hat{a}  \rangle, \langle \hat{b} \hat{C} \rangle \right\}_Q = \frac{1}{i \hbar}\left\langle \left[ \hat{a}, \hat{b} \right] \hat{C} \right\rangle =_{\Sigma} 0 \ ,
\]
meaning that all Poisson--generated vector fields are tangent to $\Sigma$, as well as that the constraint functions themselves generate no flows on $\Sigma$\ at all. In other words, $\Sigma$\ naturally inherits the Poisson structure from $\Gamma_Q$.

\subsection{Semiclassical hierarchy} \label{sec:Semiclassical}

Although, strictly speaking, the quantum phase space $\Gamma_Q$\ was constructed as a linear (or, more accurately, \emph{affine}) space, we will want to think of it as a differential manifold. More precisely, we think of it as an extension of the classical phase space $\Gamma$. The latter can be identified with $M$-dimensional subspaces of $\Gamma_Q$, formed by allowing the expectation values of the Lie algebra generators to vary, while keeping all other ``coordinates'' constant. We will designate one such subspace as being ``maximally classical'': semiclassical approximation will be valid in some small neighborhood of this subspace.

A completely specified (non-distributional) classical state assigns values in accordance with
\[
\varphi(f(x_i)) = f(\varphi(x_i)) \ ,
\]
for any function $f$. So that, e.g. the value of distance squared is the square of the value of distance and so on. We can express these ``maximal classicality'' conditions for $\Gamma_Q$\ most naturally using polynomials
\begin{equation} \label{eq:max_class}
\left\langle \widehat{pol} \right\rangle = pol(\langle \hat{x}_1 \rangle,  \langle \hat{x}_2 \rangle, \ldots ) \ ,
\end{equation}
where $pol$\ is some function polynomial in the classical coordinate functions and $\widehat{pol}\in\mathcal{A}_{kin}$\ is the corresponding element of the algebra of quantum operators. We immediately run into difficulties, as there is no unique natural identification between classical polynomial functions and elements of $\mathcal{A}_{kin}$, since different choices of ordering generally yield distinct elements. For example, $\hat{x}_1 \hat{x}_2 \hat{x}_1$\ and $(\hat{x}_1^2\hat{x}_2+\hat{x}_2\hat{x}_1^2)/2$\ are both sensible\footnote{In quantum mechanics, we also typically want physical observables to correspond to self-adjoint operators, which should then be constructed out of *-invariant combinations of the basic generators. Typically (though not always) the chosen classical Poisson algebra is real, so that the the *-structure of the enveloping algebra is defined by $\hat{x}_i^* = \hat{x}_i$, and symmetric products of the generators give *-invariant elements of $\mathcal{A}_{kin}$. Hence the choice of symmetric orderings in this example and thereafter.} choices for the quantum analogue of $x_1^2x_2$, while, in general they are not the same. A brief calculation using~(\ref{eq:CCR}) yields
\[
\hat{x}_1 \hat{x}_2 \hat{x}_1 = \frac{1}{2}(\hat{x}_1^2\hat{x}_2+\hat{x}_2\hat{x}_1^2) - \frac{\hbar^2}{2} \alpha_{21}^{\ \ i} \alpha_{1i}^{\ \ j} \hat{x}_j \ .
\]
In general, the second term on the right-hand-side does not vanish, however the expectation values it yields are ``comparatively small'', since it is suppressed by the factor of $\hbar^2$. In general, for any two choices of ordering for $\widehat{pol}$, the difference between the corresponding expectation values $\left\langle \widehat{pol} \right\rangle$\ is suppressed by $\hbar$\ (for a pair of symmetric orderings, the suppression is at least $\hbar^2$\ as in the above example, since at least two re-orderings are required). For this reason, it does not matter which ordering we choose when enforcing ``maximal classicality''~(\ref{eq:max_class}), since all choices will define classical submanifolds of $\Gamma_Q$\ that lie ``close'' to each other and will therefore define the same semiclassical neighborhood.

We follow the literature on canonical effective equations~\cite{EffAc} and~\cite{EffEq2}, and identify $\widehat{pol}$\ with the totally-symmetrized polynomials in $\hat{x}_i$, we will refer to this as \emph{Weyl ordering}. In the above example this corresponds to
\[
\widehat{x_1^2x_2} :=  \frac{1}{3} \left( \hat{x}_1^2\hat{x}_2 + \hat{x}_1\hat{x}_2\hat{x}_1 + \hat{x}_2\hat{x}_1^2 \right) =: \left( \hat{x}_1^2 \hat{x}_2 \right)_{\rm Weyl} \ .
\]

The discrepancy between polynomials in expectation values $\langle {\hat{x}_i}\rangle$\ and the expectation values of symmetrized polynomials are well captured by the so-called \emph{generalized moments} of the basic generators
\begin{equation} \label{eq:moments}
\Delta(x_1^{i_1} x_2^{i_2} \ldots ) := \left\langle \left( \hat{x}_1 - \langle \hat{x}_1 \rangle \right)^{i_1} \left( \hat{x}_2 - \langle \hat{x}_2 \rangle \right)^{i_2} \ldots \right\rangle_{\rm Weyl} \ .
\end{equation}
Note that $\Delta (x_i) = \langle \hat{x}_i -\langle \hat{x}_i \rangle \rangle = 0$. These moments are essentially non-linear functions on $\Gamma_Q$. Upon careful inspection, it is clear that the expectation value of any polynomial element of $\mathcal{A}_{kin}$ can be expressed in terms of the generalized moments and the expectation values of generators $\langle \hat{x}_i \rangle$. Together they provide an alternative (to expectation value functions) coordinate basis on $\Gamma_Q$. In particular, by writing $\hat{x}_i = \left( \hat{x}_i - \langle \hat{x}_i \rangle \right) + \langle \hat{x}_i \rangle$\ and expanding the product, it is not difficult to convince oneself that
\begin{eqnarray} \label{eq:monomial_exp}
\left\langle \hat{x}_1^{i_1} \hat{x}_2^{i_2}\ldots \right\rangle_{\rm Weyl} = & & (\langle \hat{x}_1\rangle^{i_1} \langle \hat{x}_2\rangle^{i_2} \ldots )
\\ \nonumber
&+& \left. \sum_{n_1=0}^{i_1} \sum_{n_2=0}^{i_2} \cdots  \frac{1}{n_1! n_2! \ldots} \ \frac{\partial^{\sum n_j} (x_1^{i_1} x_2^{i_2} \ldots )}{\partial^{n_1}x_1 \partial^{n_2} x_2 \ldots } \right|_{x_i=\langle \hat{x}_i \rangle} \Delta(x_1^{n_1} x_2^{n_2} \ldots ) \ , 
\end{eqnarray}
where we set $\Delta(x_1^{0} x_2^{0} \ldots ) :=0$. It follows, that condition~(\ref{eq:max_class}) with Weyl--symmetrized choice for $\widehat{pol}$\ is satisfied precisely when all moments are set to zero. Therefore, $\langle \hat{x}_i \rangle$\ and $\Delta(x_1^{i_1} x_2^{i_2} \ldots )$\ can be thought of as the classical and quantum coordinates on $\Gamma_Q$\ respectively; we will refer to this particular choice of coordinate functions on $\Gamma_Q$\ as \emph{quantum variables}. We will refer to the sum $\sum_{n=1}^M i_n$\ as the \emph{order} of the generalized moment $\Delta(x_1^{i_1} x_2^{i_2} \ldots )$. A \emph{semiclassical} state is one that is ``sharply peaked'' about some values of a classically-complete set of observables, meaning, for example, that the standard deviations of these observables are small in such a state. For our purposes, a state on $\mathcal{A}_{kin}$\ is semiclassical, or, as we will more commonly say, the corresponding point lies in the \emph{semiclassical region} of $\Gamma_Q$, if the value it assigns to all moments of order $N$\ is comparable to the value of $\hbar^{N/2}$. This comparison assumes that all classical coordinates have been rescaled so that they have the same units as $\sqrt{\hbar}$, and can be explicitly realized in ordinary quantum mechanics using Gaussian states.

An entirely equivalent definition of the semiclassical region would be obtained if we had chosen a different ordering when defining $\widehat{pol}$\ and moments in~(\ref{eq:moments}), since, because of the CCRs~(\ref{eq:CCR}) differently ordered products are equivalent up to terms proportional to powers of $\hbar$. We can therefore treat the semiclassical region of $\Gamma_Q$\ as well-defined by the choice of the generators $\hat{x}_i$\ alone and independent from our decision to work with the totally symmetrized products.

When treating a physical quantum system, in a state that lies deep within the semiclassical region, one can safely neglect high-order moments that are suppressed by many powers of $\hbar$. In this situation, one can work with a finite-dimensional \emph{truncated} system, where all moments beyond a certain order have been discarded. This is the original motivation for studying the semiclassical hierarchy and defining and studying truncated systems.

\subsection{Consistent truncation}\label{sec:Consistent_Trunc}

We define $\Gamma_Q^{(N)}$, the truncated order $N$\ semiclassical state-space of the Lie algebra $\mathfrak{a}$, as the space with coordinates $\langle \hat{x}_i \rangle$\ and $\Delta(x_1^{i_1} x_2^{i_2} \ldots )$\ up to order $N$. In essence, we will identify  $\Gamma_Q^{(N)}$ with the submanifold of $\Gamma_Q$\ defined by setting all moments beyond order $N$\ to zero. However, the quantum Poisson structure~(\ref{eq:QPoisson}) and the constraint conditions~(\ref{eq:CPoly}) need to be projected to $\Gamma_Q^{(N)}$\ with some care. Let us first define a more general notion of a semiclassical order:
\begin{eqnarray}
&(i)&  \ {\rm Order } \left( \langle \hat{x}_i \rangle \right) = 0 \ , \label{eq:orderI} \\
&(ii)& \ {\rm Order\, } \left( \Delta(x_1^{i_1} x_2^{i_2} \ldots ) \right) = \sum_{n=1}^M i_n \ , \label{eq:orderII}\\
&(iii)& \ {\rm Order\, } \left( \hbar^{n} \right) = 2n \ , \label{eq:orderIII}
\end{eqnarray}
which we extend to monomials in basic expectation values, generalized moments, and $\hbar$\ by
\[
{\rm Order\, } \left(f g \right) = {\rm Order\, } \left( f \right) + {\rm Order\, } \left( g \right) .
\]
A polynomial in $\hbar$\ and quantum variables of course generally mixes terms of different orders and does not itself posses a well--defined semicalssical order in the above sense however each of its monomial terms does. The notion of the order can be extended to polynomials as the \emph{leading semiclassical order} by defining Order $(.)$\ for a polynomial in $\hbar$\ and quantum variables as the lowest order of its non-zero monomial terms. It follows that
\begin{equation*}
{\rm Order\, } \left(f + g \right) \geq \inf \left\{ {\rm Order\, } \left( f \right), \, {\rm Order\, } \left( g \right) \right\} \ .
\end{equation*}
We define $N$--th order truncation operation on a monomial in the natural way
\begin{equation} \label{eq:Trunc1}
{\rm Trunc}_{(N)} (f) = \left\{ \begin{array}{l} f \ , \ {\rm if \ }  {\rm Order\, } (f) \leq N \ , \\ 0 \ , \ {\rm if \ }  {\rm Order\, } (f) > N \ . \end{array} \right.
\end{equation}
We extend it to polynomials by demanding linearity
\begin{equation} \label{eq:Trunc2}
{\rm Trunc}_{(N)} (f + g) = {\rm Trunc}_{(N)} (f) +{\rm Trunc}_{(N)} (g) \ .
\end{equation}
The truncation then simply cuts off all the terms of semiclassical order larger than $N$\ in the polynomial sum. The quantum Poisson bracket on $\Gamma_Q$\ consistently descends to a Lie bracket between a pair of order--$N$ truncated polynomials in a natural way
\begin{equation} \label{eq:TruncatedPB}
\{ f, g \}_Q^{(N)} := {\rm Trunc}_{(N)} \left( \{ f, g\}_Q \right) \ .
\end{equation}
In section~\ref{sec:PBOrder} we prove that
\begin{equation} \label{eq:TruncPBconsistency}
\{ f, g \}_Q^{(N)} = \{{\rm Trunc}_{(N)}( f) ,  \, {\rm Trunc}_{(N)}(g) \}_Q^{(N)} \ .
\end{equation}
ensuring that the order $N$\ truncation is a Lie algebra homomorphism from polynomials in $\hbar$\ and quantum variables of arbitrary order, to those of order $N$\ and below.
This result also has an important practical implication for using a truncated description of a dynamical quantum system. In particular, this implies that the truncated quantum Hamiltonian, ${\rm Trunc}_{(N)}( H_Q)$, can be used to compute the order $N$\ evolution of the truncated system, considerably simplifying the computations involved in equation~(\ref{eq:Qevolution}). As we explicitly prove in section~\ref{sec:PBOrder} the above consistency condition is a special consequence of a general relation  true for a pair of polynomials
\begin{equation} \label{eq:PBorder}
{\rm Order\, } \left(\{f,  g\}_Q \right) \geq {\rm Order\, } \left( f \right) + {\rm Order\, } \left( g \right) -2 \ .
\end{equation}
Unfortunately, the truncation operation is not compatible with the ordinary abelian multiplication of polynomials, since in general
\[
{\rm Trunc}_{(N)}( fg) \neq {\rm Trunc}_{(N)}( f){\rm Trunc}_{(N)}( g) \ .
\]
For example, $ {\rm Trunc}_{(3)} \left( \hbar \left\langle \left( \hat{x}_1 -  \langle \hat{x}_1 \rangle \right)^2 \right\rangle \right) = 0$, while, since both factors are of order 2
\[
{\rm Trunc}_{(3)}\left( \hbar \right)  {\rm Trunc}_{(3)} \left( \left\langle \left( \hat{x}_1 -  \langle \hat{x}_1 \rangle \right)^2 \right\rangle\right) = \hbar \left\langle \left( \hat{x}_1 -  \langle \hat{x}_1 \rangle \right)^2 \right\rangle \ .
\]
It follows that the truncated bracket defined by~(\ref{eq:TruncatedPB}) does not satisfy the Leibniz rule and is therefore not a true Poisson bracket.\footnote{The situation could be remedied by defining a truncated abelian product between polynomials along the lines of~(\ref{eq:TruncatedPB}).}

With the degrees of freedom and the Poisson structure consistently truncated to order $N$, we are left with truncating the countably infinite set of Casimir constraint conditions given e.g. by equation~(\ref{eq:CPoly}). To proceed we first express the constraints using the expectation values of basic operators and their generalized moments, which is in principle always possible---one can imagine first symmetrically reordering polynomial terms within $\left(\hat{x}_1^{i_1}\hat{x}_2^{i_2} \ldots \hat{x}_M^{i_M}\hat{C}\right)$\ and then using the expansion in equation~(\ref{eq:monomial_exp}). At this stage, it seems that applying the truncation operation of~(\ref{eq:Trunc1}) and~(\ref{eq:Trunc2}) to the constraint conditions directly will do the trick, however this leads to inconsistent results~\cite{EffCons} and generally to an incorrect reduction in the degrees of freedom as we will see in detail in section~\ref{sec:Constraints}. Instead, following~\cite{EffCons} and~\cite{EffConsRel} we formally assign the ``classical\rq{}\rq{} constraint function evaluated on the expectation values of the generators  $C(\langle \hat{x}_1 \rangle, \langle \hat{x}_2 \rangle , \ldots , \langle \hat{x}_M \rangle)$\ semiclassical order of 2 (i.e. same as $\hbar$) even though the expression involves no generalized moments or explicit factors of $\hbar$. While ultimately this particular order assignment is required for consistency, it can be physically motivated by noting that in the classical limit $C$\ must vanish on states that satisfy the Casimir constraint. In other words, in addition to the conditions~(\ref{eq:orderI})--(\ref{eq:orderIII}), when truncating the system of constraints we impose
\begin{equation} \label{eq:orderIV}
(iv) \ \  \ {\rm Order } \left( C(\langle \hat{x}_1 \rangle, \langle \hat{x}_2 \rangle , \ldots , \langle \hat{x}_M \rangle) \right) = 2 \ .
\end{equation}

In what way is the resulting truncated system of constraints \emph{consistent}? As is shown explicitly in section~\ref{sec:Constraints}, following this method of truncation we generate a finite number of non-trivial constraint conditions at each order. Moreover, provided that the non-trivial truncated constraints are functionally independent, the number of free degrees of freedom left in $\Gamma_Q^{(N)}$\ is the same as that of a system with no constraints that has one fewer basic generator. That is, provided certain regularity conditions are satisfied, the truncated tower of constraint conditions on expectation values and moments removes exactly one generating degree of freedom at each order. This is then our order $N$\ truncated phase space of the Lie algebra $\mathfrak{a}$:
\begin{quote}
The subspace $\Sigma^{(N)}$\ of the truncated space $\Gamma_Q^{(N)}$\ defined by setting the truncated set of constraint conditions $\{ {\rm Trunc}_{(N)} \left( \langle \hat{a} \hat{C} \rangle \right) : \hat{a} \in \mathcal{A}_{kin} \}$\ to zero, and equipped with a Lie bracket between polynomials in quantum variables given by~(\ref{eq:TruncatedPB}). The latter can be used to study truncated dynamics generated by a Hamiltonian or the truncated action of a symmetry group.
\end{quote}

In this section we formally defined the semiclassical truncation of states on a Lie algebra with a single center--generating (Casimir) element, generalizing the semiclassical states on the canonical algebra studied in~\cite{EffAc} and~\cite{EffEq2}. This setup has already been employed to study an $sl(2, \mathds{C})$--based quantum cosmological model in~\cite{Casimir} and we anticipate further applications to the development of effective formalism for gauge field theories where degrees of freedom at each spatial point are typically captured by a Lie algebra. In the following section, we cast the discussion of semiclassical hierarchy in algebraic form and use it to prove property~(\ref{eq:PBorder}) and the consistency result~(\ref{eq:TruncPBconsistency}). In section~\ref{sec:Constraints} we use this construction to find the conditions under which the truncated system of constraints correctly reduces the number of degrees of freedom.

\section{Extended algebra and the quantum Poisson bracket} \label{sec:PBOrder}

It is possible to define semiclassical hierarchy directly at the algebraic level rather than on the space of states, however the universal enveloping algebra $\mathcal{A}_{kin}$\ has to be extended for this purpose in order to accommodate the moment--generatig elements. We first define the \emph{classical} polynomial algebra corresponding to the Lie algebra $\mathfrak{a}$ as the commutative algebra of complex polynomials in the basic generators: $\mathcal{A}_{class} := \mathds{C}[x_1, x_2, \ldots x_M]$. When Passing from the algebraic picture to states, we will identify these ``classical variables\rq{}\rq{} $x_i$\ with the expectation values of the generators $\langle \hat{x}_i\rangle$. The extended quantum algebra is then linearly generated by finite sums of elements of $\mathcal{A}_{kin}$\ with multiplicative coefficients allowed to take values in $\mathcal{A}_{class}$. In other words $\mathcal{A}_{ext} := \mathcal{A}_{kin} \otimes \mathcal{A}_{class}\ $, and it can be algebraically generated by complex polynomials in elements of the form $\hat{x}_i$, $x_i\hat{\mathbf{1}}$, and $\hbar \hat{\mathbf{1}}$. With products between $\hat{x}_i$\ and $\hat{x}_j$\ governed by the CCRs while those between $x_i$\ and $x_j$\ as well as those between $\hat{x}_i$\ and $x_{j}$\ abelian (i.e. commutative). To emphasize that the associative product of $\mathcal{A}_{ext}$\ is non--commutative, we will use ``hats\rq{}\rq{} to denote its generic elements, $\hat{a} \in \mathcal{A}_{ext}$, just as we did with $\mathcal{A}_{kin}$.

Of particular interest to the study of semiclassical states are elements of the form $\hat{x}_i - x_i\hat{\mathbf{1}}$, which will serve as the algebraic analogue of the generalized moments. We define
\[
\widehat{\Delta x}_i :=\hat{x}_i - x_i\hat{\mathbf{1}} \ .
\]
Evidently, since $\hat{x}_i = \widehat{\Delta x}_i + x_i\hat{\mathbf{1}}$, the elements $\widehat{\Delta x}_i$\ form an alternative set of generators for $\mathcal{A}_{ext}$\ over $\mathcal{A}_{class}$. The extended algebra does possess a natural semiclassical order. As before, let us first define it for the preferred set of generators:
\begin{eqnarray}
&(i)&  \ {\rm Order } \left( x_i \hat{\mathbf{1}} \right) = 0 \ , \label{eq:AorderI} \\
&(ii)& \ {\rm Order\, } \left( \widehat{\Delta x}_i \right) =1 \ , \label{eq:AorderII}\\
&(iii)& \ {\rm Order\, } \left( \hbar \hat{\mathbf{1}}  \right) = 2 \ . \label{eq:AorderIII}
\end{eqnarray}
Once again, we extend the definition to monomials via
\begin{equation}
{\rm Order\, } \left(\hat{a} \hat{b} \right) = {\rm Order\, } ( \hat{a} ) + {\rm Order\, } ( \hat{b} )\ . \label{eq:OrderP1}
\end{equation}
In order to extend the definition of order to polynomials, we need to consider the effect of the 
 CCRs~(\ref{eq:CCR}). Distinct monomials are not all linearly independent from each other, for example, $\widehat{\Delta x}_i \widehat{\Delta x}_j$\ is assigned Order $=2$\ by above definition, while, using CCR we can express this product as
\begin{eqnarray}
\widehat{\Delta x}_i \widehat{\Delta x}_j &=& \widehat{\Delta x}_j \widehat{\Delta x}_i + \left[ \hat{x}_i - x_i\hat{\mathbf{1}}, \ \hat{x}_j - x_j\hat{\mathbf{1}} \right] \nonumber
\\ \label{eq:Reordering}
&=&  \widehat{\Delta x}_j \widehat{\Delta x}_i + \left[ \hat{x}_i, \ \hat{x}_j \right]
\\
&=& \widehat{\Delta x}_j \widehat{\Delta x}_i + i\hbar \alpha_{ij}^{\ \ k} \hat{x}_k  = \widehat{\Delta x}_j \widehat{\Delta x}_i + i\hbar \alpha_{ij}^{\ \ k} x_k \hat{\mathbf{1}}+ i\hbar \alpha_{ij}^{\ \ k} \widehat{\Delta x}_k  \ , \nonumber
\end{eqnarray} 
where the monomial terms in the final expression have orders 2, 2, and 3 respectively. We can make the definition consistent by assigning complex polynomials in $ x_i \hat{\mathbf{1}}$, $\widehat{\Delta x}_i$, and $\hbar \hat{\mathbf{1}}$\ the lowest order of all of the orders of its non-zero monomial terms expanded in any given basis (i.e. a particular ordering is chosen for the factors $\widehat{\Delta x}_i$). The above Order (.) operation on $\mathcal{A}_{ext}$\ can then be interpreted as the \emph{leading order} of a polynomial. Generalizing~(\ref{eq:Reordering}), two differently ordered monomials are equivalent up to adding terms of the same semiclassical order or higher, it follows that every element of $\mathcal{A}_{ext}$\ has a unique semiclassical order in the above sense. Some useful properties follow
\begin{eqnarray}
{\rm Order\, } \left(\hat{a} \hat{b} \right) &=& {\rm Order\, } \left( \hat{b}  \hat{a} \right) \ , \label{eq:OrderP2}\\
{\rm Order\, } \left(\hat{a} + \hat{b} \right) &\geq& \inf \left\{ {\rm Order\, } ( \hat{a} ),   \, {\rm Order\, } ( \hat{b} ) \right\} \ . \label{eq:OrderP3}
\end{eqnarray}
From~(\ref{eq:OrderP1}), (\ref{eq:OrderP2}) and~(\ref{eq:OrderP3}) it quickly follows that
\begin{equation}
{\rm Order\, } \left( \left[\hat{a}, \hat{b} \right] \right) \geq {\rm Order\, } ( \hat{a} ) + {\rm Order\, } ( \hat{b} ) \ . \label{eq:OrderP4}
\end{equation}
By virtue of the factor of $\hbar$\ appearing in the CCRs the commutator is also an ``order raising'' operation. This can be seen by considering what happens to the generators of $\mathcal{A}_{ext}$\ when commutators are taken: $[\hat{x}_i, \hbar\hat{\mathbf{1}} ] =[\hat{x}_i, x_j \hat{\mathbf{1}} ] = 0$, while $\left[ \hat{x}_i, \widehat{\Delta x}_j \right] =  i\hbar \alpha_{ij}^{\ \ k} x_k \hat{\mathbf{1}} + i\hbar \alpha_{ij}^{\ \ k} \widehat{\Delta x}_k$. In each case the order is raised by at least one relative to the second argument of the commutator. By the property of the commutator of products it quickly follows that
\begin{equation}
{\rm Order\, } \left( \left[\hat{x}_i, \hat{a} \right] \right) \geq {\rm Order\, } ( \hat{a} ) + 1 \ ,
\end{equation}
and, more generally, that
\begin{equation}
{\rm Order\, } \left( \left[\hat{a}, \hat{b} \right] \right) \geq \sup \left\{ {\rm Order\, } ( \hat{a} ),   \, {\rm Order\, } ( \hat{b} ) \right\} + 1 \ . \label{eq:OrderP5}
\end{equation}
Typically~(\ref{eq:OrderP4}) provides a stronger lower bound on the semiclassical order of a commutator $[\hat{a}, \hat{b}]$. However if the order of either $\hat{a}$\ or $\hat{b}$\ is zero, then condition~(\ref{eq:OrderP5}) is stronger.

We now establish the connection between the extended algebra and functions on the quantum phase space by defining the map $\langle . \rangle: \mathcal{A}_{ext} \rightarrow \mathscr{F}\left( \Gamma_Q\right)$, such that for any $\hat{a} \in \mathcal{A}_{kin}$\  and $f \in \mathcal{A}_{class}$ 
\begin{eqnarray} \label{eq:ExpVal}
\left\langle f(x_1, x_2, \ldots) \hat{a} \right\rangle := f(\langle \hat{x}_1 \rangle, \langle \hat{x}_2, \rangle \ldots ) \langle \hat{a} \rangle \ ,
\end{eqnarray}
where the $\langle . \rangle$\ operation on the right is performed as in Section~\ref{sec:QPhaseSpace}. The map extends to the rest of $\mathcal{A}_{ext}$\ by requiring it to be complex--linear and has a non-trivial kernel with $\langle \hat{a} \rangle = 0$\ for any element of the form $\hat{a} = \hbar^n f \widehat{\Delta x}_i$\ and linear combinations of such elements, since  $\left\langle \widehat{\Delta x}_i \right \rangle = \langle \hat{x}_i \rangle - \langle x_i \rangle =  0$. This map is compatible with the definition of the leading semiclassical order in both spaces. That is
\begin{equation} \label{eq:OrderPres}
{\rm Order }\ \left( \langle \hat{f} \rangle \right) \geq {\rm Order }\ \left( \hat{f} \right) \ .
\end{equation} 
To see this consider a symmetrized monomial $\hat{a} =  f \hbar^n \left( \widehat{\Delta x}_1^{n_1}  \widehat{\Delta x}_2^{n_2} \ldots \right)_{\rm Weyl}$, where $f \in \mathcal{A}_{class}$. By definition~(\ref{eq:AorderI})--(\ref{eq:AorderIII}) Order\,$(\hat{a}) = 2n + \sum n_i$. By looking at the definition of generalized moments~(\ref{eq:moments}) we see that this element is mapped to
\[
\langle \hat{a} \rangle = \left\langle   f(x_1, x_2 \ldots)  \hbar^n \left( \widehat{\Delta x}_1^{n_1}  \widehat{\Delta x}_2^{n_2} \ldots \right)_{\rm Weyl} \right\rangle =  f(\langle \hat{x}_1\rangle, \langle \hat{x}_2\rangle \ldots)  \hbar^n \Delta \left( x_1^{n_1} x_2^{n_2} \ldots \right) \ . 
\]
If $\langle \hat{a} \rangle = 0$, the expectation value is of ``infinite'' order. Otherwise, from definitions~(\ref{eq:orderI})--(\ref{eq:orderIII}) we obtain Order\,$(\langle \hat{a}\rangle ) = 2n + \sum n_i$. Both cases agree with~(\ref{eq:OrderPres}). Since any element of $\mathcal{A}_{ext}$\ can be expressed as a sum of monomial terms of this form, and due to linearity of $\langle.\rangle$\ and properties of Order(.) this extends to all elements of $\mathcal{A}_{ext}$.

The extended algebra inherits a non--trivial commutation bracket from the non--abelian associative product of $\mathcal{A}_{kin}$, determined by the CCRs. At the same time, the commutative algebra $\mathcal{A}_{class}$\ can be naturally equipped with the classical Poisson bracket ${\displaystyle \{x_i, x_j\} = \alpha_{ij}^{\ \ k} x_k}\ $. We combine these two brackets to construct an extended formula for computing the quantum Poisson bracket on $\Gamma_Q$\ that is more suitable for directly computing brackets between the generalized moments than equation~(\ref{eq:QPoisson}). We find that for $\hat{f}, \ \hat{g} \in \mathcal{A}_{ext}$
\begin{equation} \label{eq:QPoisson2}
\left\{ \langle \hat{f} \rangle,\langle \hat{g} \rangle \right\}_Q = \frac{1}{i\hbar} \left\langle \left[ \hat{f}, \hat{g} \right] \right\rangle + \left\langle \frac{\partial \hat{f}}{\partial x_i} \right\rangle  \left\langle \frac{\partial \hat{g}}{\partial x_j} \right\rangle \{x_i, x_j\} +    \frac{1}{i\hbar} \left\langle \frac{\partial \hat{f}}{\partial x_i} \right\rangle  \left\langle \left[ \hat{x}_i, \hat{g} \right] \right\rangle+  \frac{1}{i\hbar} \left\langle \frac{\partial \hat{g}}{\partial x_i} \right\rangle  \left\langle \left[ \hat{f}, \hat{x}_i \right] \right\rangle \ .
\end{equation}
Here by $\partial \hat{f}/\partial x_i$\ we mean the ordinary partial derivative acting on the elements of $\mathcal{A}_{class}$, so that for a monomial term, with $\hat{a} \in \mathcal{A}_{kin}$\  and $h \in \mathcal{A}_{class}$
\[
\frac{\partial (h\hat{a} )}{\partial x_i} := \frac{\partial h}{\partial x_i} \hat{a} \ .
\]
It is immediately clear that the formula agrees with the basic definition~(\ref{eq:QPoisson}), when $\hat{f}$\ and $\hat{g}$\ are both in the subalgebra $\mathcal{A}_{kin}$, since in this case only the first term is non-zero. It is also clear upon inspection that the RHS defines a bilinear map over $\mathbb{C}$. Therefore, in order to show that~(\ref{eq:QPoisson2}) holds generally, we only need to show that it holds for a pair of monomials $f\hat{a}$\ and $g \hat{b}$\ with $f, g \in \mathcal{A}_{class}$\ and $\hat{a}, \hat{b} \in \mathcal{A}_{kin}$, since a general element of $\mathcal{A}_{ext}$\ is a linear combination of such terms. Using the quantum Poisson bracket of section~\ref{sec:QPhaseSpace} we have
\begin{eqnarray*}
\left\{ \langle f\hat{a} \rangle,\langle g\hat{b} \rangle \right\}_Q
&=& \left\{ \langle f \rangle \langle \hat{a} \rangle, \langle g \rangle \langle \hat{b} \rangle \right\}_Q
\\
 &=& \langle f \rangle  \langle g \rangle \left\{  \langle \hat{a} \rangle, \langle \hat{b} \rangle \right\}_Q + \langle \hat{a} \rangle \langle \hat{b} \rangle \left\{ \langle f \rangle , \langle g \rangle  \right\}_Q + \langle g \rangle \langle \hat{a} \rangle \left\{ \langle f \rangle ,  \langle \hat{b} \rangle \right\}_Q+ \langle f \rangle \langle \hat{b} \rangle \left\{  \langle \hat{a} \rangle, \langle g \rangle  \right\}_Q
 \\
&=& \langle f \rangle  \langle g \rangle  \frac{1}{i\hbar} \left\langle \left[ \hat{a}, \hat{b} \right] \right\rangle + \langle \hat{a} \rangle \langle \hat{b} \rangle \left\langle \frac{\partial f}{\partial x_i} \right\rangle  \left\langle \frac{\partial g}{\partial x_j} \right\rangle \frac{1}{i\hbar} \left\langle \left[ \hat{x}_i, \hat{x}_j \right] \right\rangle
\\
& & + \langle g \rangle \langle \hat{a} \rangle \left\langle  \frac{\partial f}{\partial x_i} \right\rangle  \frac{1}{i\hbar} \left\langle \left[ \hat{x}_i, \hat{b} \right] \right\rangle + \langle f \rangle \langle \hat{b} \rangle  \left\langle \frac{\partial g}{\partial x_i} \right\rangle \frac{1}{i\hbar} \left\langle \left[ \hat{a}, \hat{x}_i \right] \right\rangle \ .
\end{eqnarray*}
The first equality follows straight from~(\ref{eq:ExpVal}), the second equality utilizing the product rule, which is again used to write the 2nd, 3rd, and 4th terms in the final expression. It is straightforward to see that the right-hand-side of~(\ref{eq:QPoisson2}) yields an identical result. The latter therefore holds for all elements of $\mathcal{A}_{ext}$.

We are almost in a position to evaluate the order of the terms on the right-hand side of~(\ref{eq:QPoisson2}) and to verify relation~(\ref{eq:PBorder}) using the properties of ``algebraic'' semiclassical order and the order-preserving property of $\langle . \rangle$. It only remains to understand the effect that the partial derivatives with respect to $x_i$\ have on the semiclassical order. We first note that the derivative does not \emph{reduce} the order of generators $\hbar \hat{\boldsymbol{1}}$\ and $x_i\hat{\boldsymbol{1}}$, while
\[
\frac{\partial}{\partial x_i} \widehat{\Delta x}_j = - \delta_{ij} \hat{\boldsymbol{1}} \ ,
\]
so that, acting on any one of these generators of $\mathcal{A}_{ext}$\ , $\partial/\partial x_i$\ reduces the semiclassical order by 1 at the most. In fact, there is an elegant result that is not strictly necessary for our discussion here, but can be quite useful in another context
\[
\frac{\partial}{\partial x_i} \left( \widehat{\Delta x}_1^{n_1} \ldots \widehat{\Delta x}_i^{n_i} \ldots \widehat{\Delta x}_M^{n_M} \right)_{\rm Weyl} = - n_i \left( \widehat{\Delta x}_1^{n_1} \ldots \widehat{\Delta x}_i^{n_i-1} \ldots \widehat{\Delta x}_M^{n_M} \right)_{\rm Weyl} \ .
\]
Since the derivative obeys $\partial (\hat{a} \hat{b})/\partial x_i = (\partial \hat{a} /\partial x_i) \hat{b} + \hat{a} (\partial  \hat{b}/\partial x_i )$\ , and using~(\ref{eq:OrderP1}) and~(\ref{eq:OrderP2}), we conclude that the derivative reduces the order of any monomial by 1 at the most. Further, since $\partial/\partial x_i$\ is linear, and again employing~(\ref{eq:OrderP1}) and~(\ref{eq:OrderP2}), we conclude that
\begin{equation}\label{eq:OrderP6}
{\rm Order\, } \left( \frac{\partial \hat{a} }{\partial x_i} \right) \geq {\rm Order\, } \left( \hat{a} \right) - 1 \ , \quad {\rm for \ any \ \ } \hat{a} \in \mathcal{A}_{ext} \ .
\end{equation} 
Finally, by inspection and referring to properties~(\ref{eq:OrderP1}), ~(\ref{eq:OrderP2})--(\ref{eq:OrderP4}), ~(\ref{eq:OrderP5}), (\ref{eq:OrderPres}), ~(\ref{eq:OrderP6}), we see that the semiclassical order of every term on the right-hand side of~(\ref{eq:QPoisson2}) is bounded from below by ${\rm Order\, } ( \hat{f} ) + {\rm Order\, } ( \hat{g} ) - 2$, which proves~(\ref{eq:PBorder}) for the case where $f = \langle \hat{f} \rangle$\ and $g = \langle \hat{g} \rangle$\ for some $\hat{f}, \hat{g} \in \mathcal{A}_{ext}$\ . Since all of the quantum variables can be written in this way and since they provide a complete coordinate basis on $\Gamma_Q$\ the property~(\ref{eq:PBorder})
\begin{equation*}
{\rm Order\, } \left(\{f,  g\}_Q \right) \geq {\rm Order\, } \left( f \right) + {\rm Order\, } \left( g \right) -2 \ ,
\end{equation*}
holds for any pair of polynomial functions $f$\ and $g$\ on $\Gamma_Q$\ . For $N\geq2$, this implies that polynomial functions of order $N$\ and above form a \emph{Poisson ideal}, that is, for a polynomial function $f \in \mathscr{F}(\Gamma_Q)$, such that Order $(f) \geq N$
\[
{\rm Order\, } \left( \{f, g\}_Q \right) \geq N \ , \quad {\rm for \ any \ polynomial \ function \ \ } g \in \mathscr{F}(\Gamma_Q) \ .
\]
The consistency result for the truncated quantum Poisson bracket~(\ref{eq:TruncPBconsistency}) follows immediately for $N\geq1$, since the portions of $f$\ and $g$\ removed by truncation are themselves of order $(N+1)$\ or above and would generate terms of order $(N+1)$\ and above when their quantum Poisson bracket with other functions is taken.

\section{Counting truncated constraints} \label{sec:Constraints}

In this section we count the truncated set of Casimir constraint functions generated by a single quantum constraint as in~(\ref{eq:CPoly}) order-by-order in the semiclassical expansion. The counting establishes the conditions under which the Casimir constraint removes the number of degrees of freedom equivalent to a single generator.

At this stage it is convenient to introduce some further short-hand notation. The Weyl-ordered products of the generators $\widehat{\Delta x}_i$, which form an $\mathcal{A}_{class}$--linear basis on $\mathcal{A}_{ext}$, defined in section~\ref{sec:PBOrder}, will be denoted by
\[
\hat{e}_{\vec{i}} = \hat{e}_{(i_1, i_2, \ldots i_M)} := \left(\widehat{\Delta x}_1^{i_1} \widehat{\Delta x}_2^{i_2} \ldots \widehat{\Delta x}_M^{i_M} \right)_{\rm Weyl-ordered} \ .
\]
Here $\vec{i}$\ is an $M$-tuple of non-negative integers. We will use the convention $\hat{e}_0=\hat{\mathbf{1}}$. For convenience we also define the \textbf{degree} $| \vec{i} | := \sum_{n=1}^M i_n$\ and \textbf{partial ordering} $\vec{i} \geq \vec{j}$\ if $i_n \geq j_n\ \ \forall n$, so that $\vec{i} > \vec{j}$\ if $\vec{i} \geq \vec{j}$\ and $\vec{i} \neq \vec{j}$. We will denote the expectation values of the basis elements by
\[
\varepsilon_{(i_1, i_2, \ldots i_M)} =  \varepsilon_{\vec{i}} =
\langle \hat{e}_{\vec{i}} \rangle \ .
\]
By construction $\varepsilon_{\vec{i}}=0$, for all $\vec{i}$, with $| \vec{i}| = 1$\ , and $\varepsilon_{0}=1$. We will use $\vec{i}!$\ to denote $(i_1!i_2! \ldots i_M!)$. Since one can symmetrically order an arbitrary product of $\widehat{\Delta x}_i$-s by adding terms that are of lower polynomial degree and proportional to powers of $\hbar$
\begin{eqnarray*}
\hat{e}_{\vec{i}} \hat{e}_{\vec{j}} &=& \hat{e}_{(\vec{i}+\vec{j})} + \hbar \sum_{\tiny{\begin{array}{c} \vec{k} < (\vec{i}+\vec{j})
\\
|\vec{k}| \leq |\vec{i}+\vec{j}|-1 \end{array}}} \beta_{\vec{i}, \vec{j}}^{(1) \ \vec{k}} \hat{e}_{\vec{k}} + \hbar^2 \sum_{\tiny{\begin{array}{c} \vec{k} < (\vec{i}+\vec{j})
\\
|\vec{k}| \leq |\vec{i}+\vec{j}|-2 \end{array}}} \beta_{\vec{i}, \vec{j}}^{(2) \ \vec{k}} \hat{e}_{\vec{k}} + \ldots \ ,
\end{eqnarray*}
where $\beta_{\vec{i}, \vec{j}}^{(n) \ \vec{k}}$\ are some polynomials in $x_i$.

Starting from section~\ref{sec:counting} we will predominantly work with quantum variables rather than their algebraic analogues, and it will be convenient to use $x_i$\ to denote the expectation values $\langle \hat{x}_i \rangle$\ as well as their ``classical placeholders'' in $\mathcal{A}_{ext}$. In addition, the classical polynomial for the Casimir constraint $C(x_1, x_2, \ldots x_M)$\ will feature prominently and we will denote
\[
C := C ( \langle \hat{x}_1 \rangle, \langle \hat{x}_2 \rangle \ldots \langle \hat{x}_M \rangle ) \equiv \langle C(x_1, x_2, \ldots x_M) \hat{\boldsymbol{1}} \rangle \ .
\]
where $\langle . \rangle$\ in the final expression is taken in the sense of~(\ref{eq:ExpVal}). Similarly we will denote
\[
\frac{\partial C}{\partial x_i} := \left\langle \frac{\partial C}{\partial x_i} \hat{\boldsymbol{1}}\right\rangle \ , \quad {\rm and \ so \ forth.}
\]

\subsection{Constraints and truncation} \label{sec:truncation}

As discussed in section~\ref{sec:QLieAlg}, the Casimir constraint can be imposed by demanding that for all polynomial functions $f$, we have $\langle f(\hat{x}_1, \hat{x}_2 \ldots \hat{x}_M) \hat{C} \rangle = 0$. For the purposes of truncation, these conditions can be systematically imposed by using the basis $\{\hat{e}_{\vec{i}} \}$:
\[
C_{\vec{i}} := \langle \hat{e}_{\vec{i}\ } \hat{C} \rangle = 0, \ \ \forall \ \vec{i} \in \mathbb{Z}_+^M \ .
\]
With our conventions $C_0 := \langle \hat{C} \rangle$. This is equivalent to the set of conditions given by equation~(\ref{eq:CPoly}).

There are two distinct aspects to the truncation process: 1) truncation of the degrees of freedom; 2) truncation of the above system of the constraint functions. For concreteness, assume we truncate at some order $N \geq 2$\ of the semiclassical expansion. Degrees of freedom are truncated by dropping moments of degree greater than $N$, i.e. drop all $\varepsilon_{\vec{i}}$\ that have $| \vec{i} | > N$. As briefly discussed in section~\ref{sec:Consistent_Trunc}, the truncation of the system of constraints is more subtle. $C_{\vec{i}}$\ are linear functions of the moments $\varepsilon_{\vec{j}}$\ and for the purposes of truncation, we will separate terms appearing in their expressions into three types:
\begin{itemize}
\item $f(\langle \hat{x}_1 \rangle, \langle \hat{x}_2 \rangle \ldots \langle \hat{x}_M \rangle ) \varepsilon_{\vec{i}}$, where $f$\ is a polynomial in the expectation values, is assigned semiclassical order equal to $| \vec{i} |$.
\item $C(\langle \hat{x}_1 \rangle, \langle \hat{x}_2 \rangle \ldots \langle \hat{x}_M \rangle) \varepsilon_{\vec{i}}$, where $C$\ is the classical polynomial expression for the constraint, is assigned, as an exception to the previous point, semiclassical order $\left( | \vec{i} |+ 2 \right)$.
\item $\hbar^n f(\langle \hat{x}_1 \rangle, \langle \hat{x}_2 \rangle \ldots \langle \hat{x}_M \rangle) \varepsilon_{\vec{i}}$, arises upon reordering algebra elements and is assigned semiclassical order equal to $\left( | \vec{i} |+ 2n \right)$.
\end{itemize}
Upon truncation, terms in the expressions for the constraint functions of the semiclassical order higher than $N$\ are dropped, just as prescribed by~(\ref{eq:Trunc1}) and~(\ref{eq:Trunc2}).

\subsection{Counting degrees of freedom} \label{sec:counting}

First we count the degrees of freedom of an unconstrained system
generated by polynomials in $M$\ basic variables. The system is
parameterized by $M$\ expectation values and at each semiclassical
order $N \geq 2$\ the freedoms are represented by the independent
functions $\varepsilon_{\vec{i}}$, where $| \vec{i} | = N$.
\emph{How many such variables are there?}---As many as the number of
$M$-tuples of non-negative integers with $|\vec{i}|=N$, we will
refer to this as $\mathcal{N}_M(N)$. Combinatorially, each such
$M$-tuple is produced by considering a row of $N+M-1$\ identical
objects and marking $M-1$ of them to serve as partitions, the value
of $i_n$\ is then the number of unmarked objects between partition
$(n-1)$\ and partition $n$ (where partitions $0$ and $M$\ are
assumed to be at the ends). The answer is then simply
\[
\mathcal{N}_M(N) = {N+M-1 \choose M-1} \ .
\]
Imposing the Casimir condition removes a single classical degree of freedom. We thus expect the
tower of constraint conditions to remove the equivalent of one
combinatorial degree of freedom from the algebra of observables.
I.e. after the constraints are imposed, we expect as many degrees of
freedom as for a system with $(M-1)$\ generators to remain, so that
at each semiclassical order we should then have
$\mathcal{N}_{M-1}(N) = {N+M-2 \choose M-2}$\ free variables. Using
the identity ${a\choose b} - {a-1 \choose b} = {a-1\choose b-1}$,
which is straightforward to verify, we conclude that the required
number of independent conditions at each order is ${N+M-2 \choose
M-1}$.

\emph{How many constraint conditions are there at each order?} This
may seem difficult to answer, as constraint conditions
$C_{\vec{i}}$\ generally mix terms of different orders. We will
proceed by first showing that, when truncated at a given order, the
system of constraints becomes finite. The number of constraints at
each order is then the number of additional non-trivial constraint
conditions that arise when we raise the truncation order by one.

Analogously to relation~(\ref{eq:monomial_exp}), by writing $\hat{x}_i = \widehat{\Delta x}_i + x_i\mathbf{\hat{1}}$, we can expand any symmetrized monomial as
\begin{equation}\label{eq:C_expansion}
\left( \hat{x}_1^{i_1} \hat{x}_2^{i_2} \ldots \right)_{\rm Weyl} = \sum_{\vec{j} \leq \vec{i}} \frac{1}{\vec{j}!} \frac{\partial^{|\vec{j}|} \left( x_1^{i_1} x_2^{i_2} \ldots \right)}{\partial^{j_1} x_1 \partial^{j_2} x_2 \ldots } \hat{e}_{\vec{j}} \ .
\end{equation}
This easily extends to polynomials that are sums of symmetrized monomials. Of course, any monomial can be symmetrized by adding lower polynomial order terms proportional to powers of $\hbar$. In particular, the constraint operator itself can be expressed in the form
\[
\hat{C} = \sum_{\vec{j} \leq \vec{i}} \frac{1}{\vec{j}!} \frac{\partial^{|\vec{j}|} C}{\partial^{j_1} x_1 \partial^{j_2} x_2 \ldots } \hat{e}_{\vec{j}} + \hbar \times \left( {\rm terms\ of\ order\ 0\ and\ higher} \right) \ .
\]
A general constraint function therefore has the form
\begin{eqnarray}\label{eq:constraints}
C_{\vec{i}} &=& \sum_{\vec{j} \geq \vec{i}} \frac{1}{(\vec{j} -
\vec{i})!} \frac{\partial^{|\vec{j} - \vec{i}|} C}{\partial
x_1^{j_1-i_1} \ldots \partial x_{M}^{j_M-i_M}} \varepsilon_{\vec{j}}
+ \hbar \sum_{|\vec{j}| \geq |\vec{i}|-1} \alpha_{C_{\vec{i}}}^{(1)
\ \vec{j}} \varepsilon_{\vec{j}} \nonumber \\ & & + \hbar^2
\sum_{|\vec{j}| \geq |\vec{i}|-2} \alpha_{C_{\vec{i}}}^{(2) \
\vec{j}} \varepsilon_{\vec{j}} + \ldots + \hbar^{|\vec{i}|}
\sum_{\vec{j}} \alpha_{C_{\vec{i}}}^{(|\vec{i}|) \ \vec{j}}
\varepsilon_{\vec{j}} \ .
\end{eqnarray}
Here $\alpha_{C_{\vec{i}}}^{(n) \ {\vec{j}}}$\ are coefficients
polynomial in the expectation values $x_i$, and thus of semiclassical order $0$ (not to be confused with the Lie algebra structure constants $\alpha_{ij}^{\ k}$\ of sections~\ref{sec:QLieAlg} and~\ref{sec:PBOrder}). The first sum comes from the Weyl-symmetric
part of the element $\hat{e}_{\vec{i}\ } \hat{C}$, subsequent sums
arise from its components that are antisymmetric in one, two and
more adjacent pairs of moment-generating elements $\widehat{\Delta
x}_i$: each antisymmetric pair can be reduced by using the CCRs thus
producing the powers of $\hbar$. Additional terms in the second, third and further sums come from the expectation value of the product between $\hat{e}_{\vec{i}}$\ and terms multiplied by $\hbar$\ in the expression~(\ref{eq:C_expansion}) for the constraint element: such terms are of order $(|\vec{i}|+2)$\ or higher.

The important feature of the above expansion is that the lowest
semiclassical order terms in the first sum are $C\varepsilon_{\vec{i}}$\ and $\sum_k
\frac{\partial C}{\partial x_k} \varepsilon_{(i_1, \ldots, i_k+1,
\ldots, i_M)}$. Since for the purposes of truncating the constraints
$C$\ is of order $2$, the latter term has the lowest order, which is
$(|\vec{i}|+1)$. Terms coming from the second sum are at least of the order $(|\vec{i}|+1)$, the rest of the sums contribute terms of order $(|\vec{i}|+2)$\ and above. Thus, after truncation at order $N$, constraints
$C_{\vec{i}}=0$\ are satisfied identically for all $|\vec{i}| >N-1$.
The number of non-trivial conditions up to order $N$\ is the same as
the number of non-negative integer $M$-tuples of degree $N-1$\ and
less. The change in this number as we go from truncation at order
$N-1$\ to truncation at order $N$\ is the same as the number of
$M$-tuples of degree $N-1$, namely $\mathcal{N}_M(N-1)={N+M-2
\choose M-1}$\ as required by the counting.

This shows that, provided the non-trivial constraint conditions
remaining after truncation are functionally independent, they remove
precisely one combinatorial degree of freedom. In the next section,
we prove that under a reasonable set of conditions, which is
sufficiently broad for our purposes, the truncated set of
constraints is indeed functionally independent.

\subsection{Independence of truncated constraints}

In this section we look for conditions under which the set of
constraint functions $\{C_{\vec{i}}\}$\ truncated at some order $N$\
in accordance with Section~\ref{sec:truncation} is \emph{functionally
independent}. We first formulate the conditions needed for the correct
reduction of degrees of freedom. Then, using semiclassical orders, we
argue that a simpler set of conditions, more amenable to direct
analysis, is sufficient for semiclassical states. Detailed analysis
is carried out in subsection~\ref{sec:theorems}.\\

In general, we consider a (finite) set of functions $\{f_i\}_{i=1,
\ldots L}$\ \emph{functionally} independent in some region if fixing
the values of all these functions defines a hypersurface of
codimension $L$. That is, locally these functions fix $L$\ degrees
of freedom---precisely the desired property for our system of
constraints. We can apply a version of the Frobenius Integrability
Theorem to the set of exterior derivatives $\{ df_i \}_{i=1, \ldots
L}$: the functions integrate to a hypersurface of codimension $L$\
if and only if the set of their exterior derivatives is
\emph{linearly} independent at every point in the region of
interest.

We assume that in the original classical system the single (classical) constraint removes a
single classical degree of freedom. By the integrability
considerations, this implies that the constraint polynomial must be regular on
the constraint surface, i.e. $dC|_{\tiny{C=0}} \neq 0$. By
continuity, this must also hold in some neighborhood of $C=0$.

Thus, for correct reduction of the degrees of freedom, we need to
show that $\{d_QC_{\vec{j}} \}_{|\vec{i}|\leq N-1}$\ are linearly
independent. Here we treat the quantum phase space $\Gamma_Q$, prior
to imposing the constraint, as the cartesian product
$\Gamma_Q=\Gamma \times \mathcal{M}_{\varepsilon}$ of the classical
phase space $\Gamma$\ and the space of moments
$\mathcal{M}_{\varepsilon}$. The exterior derivative on $\Gamma_Q$\
is the (direct) sum of exterior derivatives on the component spaces
$d_Q = d + d_{\varepsilon}$. The covariant coordinate vectors are
$d_Qx_i = dx_i$\ and $d_Q\varepsilon_{\vec{i}} = d_{\varepsilon}
\varepsilon_{\vec{i}}$. Naturally a pair of non-zero gradients $df$\ and
$d_{\varepsilon}g$\ are always linearly independent as they belong
to different disjoint subspaces of the cotangent space. In
analyzing semiclassical orders below we treat all non-truncated
coordinate directions on equal footing. Thus the statement ``$d_Qf$\ has
leading contribution of order $n$'' will mean that the
\emph{coefficient} for one of the coordinate covectors $dx_i$\ or
$d_{\varepsilon} \varepsilon_{\vec{i}}$\ in the coordinate
decomposition of $d_Qf$\ is of semiclassical order $n$, while others
have coefficients of equal or higher order.

Let us specialize the expression for the general constraint function $C_{\vec{i}}$~(\ref{eq:constraints}) to truncation at order $N$\ as
described in Section~\ref{sec:truncation}. For $|\vec{i}| < N-1$\
the sums over $\vec{j}$\ terminate when $|\vec{j}|=N-2n$, where $n$\
is the power of $\hbar$, multiplying the sum. In particular,
\begin{eqnarray}\label{eq:trnc_constraints}
C_{\vec{i}} &=& \sum_{\tiny{\begin{array}{c} \vec{j} \geq \vec{i} \\
1< |\vec{j}| \leq N \end{array}}} \frac{1}{(\vec{j} - \vec{i})!}
\frac{\partial^{|\vec{j} - \vec{i}|} C}{\partial x_1^{j_1-i_1}
\ldots \partial x_{M}^{j_M-i_M}} \varepsilon_{\vec{j}} \ + \ \hbar
\sum_{N-2\geq |\vec{j}| \geq |\vec{i}|-1} \alpha_{C_{\vec{i}}}^{(1)
\ \vec{j}} \varepsilon_{\vec{j}} \nn\\ && + \ \ \hbar^2\,\,
\sum_{N-4 \geq |\vec{j}| \geq |\vec{i}|-2} \alpha_{C_{\vec{i}}}^{(2)
\ \vec{j}} \varepsilon_{\vec{j}} + \ldots \ .
\end{eqnarray}
For $|\vec{i}| = N-1$\ the sums terminate in the same way, however
the first term in the expansion $C\varepsilon_{\vec{i}}$\ is treated
as having semiclassical order $|\vec{i}|+2 = N+1 >N$\ and thus is
dropped upon truncation.
\begin{eqnarray}\label{eq:trnc_constraints2}
C_{\vec{i}} &=& \sum_{\tiny{\begin{array}{c} \vec{j} > \vec{i} \\
1< |\vec{j}| \leq N \end{array}}} \frac{1}{(\vec{j} - \vec{i})!}
\frac{\partial^{|\vec{j} - \vec{i}|} C}{\partial x_1^{j_1-i_1}
\ldots \partial x_{M}^{j_M-i_M}} \varepsilon_{\vec{j}} + \hbar
\sum_{|\vec{j}|=N-2} \alpha_{C_{\vec{i}}}^{(1) \ \vec{j}}
\varepsilon_{\vec{j}} \nn\\ &=& \sum_{l=1}^{M} \frac{\partial
C}{\partial x_l} \varepsilon_{(i_1, \ldots, i_l+1, \ldots)} + \hbar
\sum_{|\vec{j}|=N-2} \alpha_{C_{\vec{i}}}^{(1) \ \vec{j}}
\varepsilon_{\vec{j}} \ .
\end{eqnarray}
Higher reordering terms get dropped as they are of order higher than
$N$.

From~(\ref{eq:trnc_constraints}) and~(\ref{eq:trnc_constraints2}) we
infer the following qualitative features of $d_QC_{\vec{i}} =
dC_{\vec{i}} + d_{\varepsilon}C_{\vec{i}}$:
\begin{itemize}
\item leading order contribution to $dC_0$\ comes from $dC$\ and is
of semiclassical order $0$
\item leading order contribution to all other $dC_{\vec{i}}$\ is at least of
semiclassical order $2$
\item contributions to $d_{\varepsilon} C_{\vec{i}}$\ start at order
$0$.
\end{itemize}
We infer, that to leading semiclassical order, $d_QC_0$\ is linearly
independent of $\{ d_Q C_{\vec{i}}\}_{1\leq|\vec{i}|\leq N-1}$.
Furthermore, to establish linear independence of $\{d_Q
C_{\vec{i}}\}_{1\leq|\vec{i}|\leq N-1}$\ to leading semiclassical
order we only need to concern ourselves with the linear independence
of $\{ d_{\varepsilon}C_{\vec{i}} \}_{1\leq |\vec{i}|\leq N-1}$.

Taking gradient of the truncated constraint functions with respect
to the moments we obtain, for $|\vec{i}| < N-1$:
\begin{eqnarray*}
d_{\varepsilon}C_{\vec{i}} &=& \sum_{\tiny{\begin{array}{c} \vec{j} \geq \vec{i} \\
1<|\vec{j}| \leq N \end{array}}} \frac{1}{(\vec{j} - \vec{i})!}
\frac{\partial^{|\vec{j} - \vec{i}|} C}{\partial x_1^{j_1-i_1}
\ldots \partial x_{M}^{j_M-i_M}} d_{\varepsilon}
\varepsilon_{\vec{j}} + \hbar \sum_{N-2 \geq |\vec{j}| \geq
|\vec{i}|-1} \alpha_{C_{\vec{i}}}^{(1) \ \vec{j}} d_{\varepsilon}
\varepsilon_{\vec{j}}\\ &&+ \hbar^2 \sum_{N-4 \geq |\vec{j}| \geq
|\vec{i}|-2} \alpha_{C_{\vec{i}}}^{(2) \ \vec{j}} d_{\varepsilon}
\varepsilon_{\vec{j}} + \ldots \ .
\end{eqnarray*}
A similar expression follows for $|\vec{i}| = N-1$\ by
using~(\ref{eq:trnc_constraints2}). In either case, the leading
order contribution to the gradient along each direction
$d_{\varepsilon} \varepsilon_{\vec{i}}$, is given by the terms in
the first sum alone, other contributions are suppressed by powers of
$\hbar$. We will denote this `symmetric' part of $C_{\vec{i}}$\ by
$\tilde{C}_{\vec{i}}$\ so that  for $|\vec{i}| < N-1$:
\begin{eqnarray*}
d_{\varepsilon}\tilde{C}_{\vec{i}} = \sum_{\tiny{\begin{array}{c} \vec{j} \geq \vec{i} \\
1 < |\vec{j}| \leq N \end{array}}} \frac{1}{(\vec{j} - \vec{i})!}
\frac{\partial^{|\vec{j} - \vec{i}|} C}{\partial x_1^{j_1-i_1}
\ldots \partial x_{M}^{j_M-i_M}} d_{\varepsilon}
\varepsilon_{\vec{j}} \ .
\end{eqnarray*}
A similar expression follows for $|\vec{i}| = N-1$. As all the
constraint functions are linear in the moments
$\varepsilon_{\vec{i}}$, the gradients $d_{\varepsilon}C_{\vec{i}}$\
as well as their symmetric parts $d_{\varepsilon}
\tilde{C}_{\vec{i}}$\ have coefficients that depend on the
expectation values alone, and are
thus much easier to analyze than $d_QC_{\vec{i}}$.\\

In the next subsection we focus on linear independence of $\{
d_{\varepsilon}\tilde{C}_{\vec{i}} \}_{1\leq |\vec{i}|\leq N-1}$\
and briefly touch upon the way higher order corrections may affect
the results. From the results established below we conclude, that
for a sufficiently semiclassical state:
\begin{itemize}\item for expectation values satisfying the
classical constraint function, the truncated quantum constraints are
functionally independent so long as $dC$\ is not comparable to
$\hbar$\ or the moments in at least one coordinate direction

\item for expectation values off the classical constraint surface,
the constraint functions are functionally independent so long as for
some $k$\ neither $\frac{\partial C}{\partial x_k}$\ nor
$\frac{\partial^{N-2}}{\partial x_k^{N-2}}\left( \frac{1}{C}
\right)$\ are comparable to $\hbar$\ or the moments.

\end{itemize}

While these conditions can be violated, this is likely to happen
sufficiently ``far away'' from the classical constraint surface as
near $C=0$, $\frac{1}{C}$\ and its derivatives blow up. In addition,
these conditions are \emph{sufficient}, but \emph{not necessary} and
in some cases the constraints may be independent even if they do not
hold. In any event, the above conditions provide a viable and rigid
test of whether the semiclassical truncation of a given constrained system
reduces the degrees of freedom correctly.

\subsection{Details of the argument} \label{sec:theorems}

\textbf{Key result}: \emph{for the system of constraints truncated
at order $N$, the set of gradients $\{
d_{\varepsilon}\tilde{C}_{\vec{i}} \}_{1\leq |\vec{i}|\leq N-1}$\ is
linearly independent when expectation values $x_i$\ lie in some
neighborhood of the classical constraint surface $C=0$.} (Note: this
does not place any conditions on the values of moments.)\\ \ \\

\textbf{Proof}: We need only prove that the gradients are linearly
independent when the expectation values \emph{do} lie on the
classical constraint surface. By continuity, they remain independent
in some open neighborhood of the surface $C=0$.\\ \ \\

Suppose that the set $\{ d_{\varepsilon}\tilde{C}_{\vec{i}}
\}_{1\leq |\vec{i}|\leq N-1}$\ is linearly dependent at some point
$P\in \Gamma_Q$, then there are numerical coefficients
(with appropriate units) $\gamma^{\vec{i}}$\ such that
\[
\sum_{1\leq|\vec{i}|\leq N-1} \left. \gamma^{\vec{i}}
d_{\varepsilon} \tilde{C}_{\vec{i}} \right|_{\tiny{P}} = 0 \ .
\]
However,
\begin{eqnarray*}
\sum_{1\leq|\vec{i}|\leq N-1} \left.
\gamma^{\vec{i}}d_{\varepsilon}\tilde{C}_{\vec{i}}
\right|_{\tiny{P}} &=& \sum_{1\leq|\vec{i}| < N-1}
\sum_{\tiny{\begin{array}{c} \vec{j} \geq \vec{i}\\ 1<|\vec{j}| \leq
N \end{array}}} \left. \gamma^{\vec{i}} \frac{1}{(\vec{j} -
\vec{i})!} \frac{\partial^{|\vec{j} - \vec{i}|} C}{\partial
x_1^{j_1-i_1} \ldots \partial x_{M}^{j_M-i_M}} d_{\varepsilon}
\varepsilon_{\vec{j}} \right|_{\tiny{P}} \\ && +  \sum_{|\vec{i}| =
N-1} \sum_{\tiny{\begin{array}{c} \vec{j} > \vec{i}\\ |\vec{j}| = N
\end{array}}} \left. \gamma^{\vec{i}} \frac{1}{(\vec{j} - \vec{i})!}
\frac{\partial^{|\vec{j} - \vec{i}|} C}{\partial x_1^{j_1-i_1}
\ldots \partial x_{M}^{j_M-i_M}} d_{\varepsilon}
\varepsilon_{\vec{j}} \right|_{\tiny{P}} \ .
\end{eqnarray*}
Two separate (double) sums appear due to the subtlety of truncation,
since the terms $C\varepsilon_{\vec{j}}$\ are dropped for
$|\vec{j}|\geq N-1$, so that the terms with $\vec{j}=\vec{i}$\ do not appear for $|\vec{i}|=N-1$. Notice that in the first (double)
sum above $|\vec{i}| < N-1$\ so that, when $|\vec{j}|=N-1$\ or $|\vec{j}|=N$, we always have $\vec{j}>\vec{i}$. We combine the $|\vec{j}|=N-1$\ and $|\vec{j}|=N$\ terms from the first double sum with the second double sum to obtain
\begin{eqnarray*}
\sum_{1\leq|\vec{i}|\leq N-1} \left.
\gamma^{\vec{i}}d_{\varepsilon}\tilde{C}_{\vec{i}}
\right|_{\tiny{P}} &=& \sum_{1\leq|\vec{i}| < N-1}
\sum_{\tiny{\begin{array}{c} \vec{j} \geq \vec{i}\\ 1<|\vec{j}| <
N-1 \end{array}}} \left. \gamma^{\vec{i}} \frac{1}{(\vec{j} -
\vec{i})!} \frac{\partial^{|\vec{j} - \vec{i}|} C}{\partial
x_1^{j_1-i_1} \ldots \partial x_{M}^{j_M-i_M}} d_{\varepsilon}
\varepsilon_{\vec{j}} \right|_{\tiny{P}}
\\
&& +  \sum_{1 \leq |\vec{i}| \leq N-1} \sum_{\tiny{\begin{array}{c} \vec{j} > \vec{i}\\ N-1 \leq |\vec{j}| \leq N
\end{array}}} \left. \gamma^{\vec{i}} \frac{1}{(\vec{j} - \vec{i})!}
\frac{\partial^{|\vec{j} - \vec{i}|} C}{\partial x_1^{j_1-i_1}
\ldots \partial x_{M}^{j_M-i_M}} d_{\varepsilon}
\varepsilon_{\vec{j}} \right|_{\tiny{P}} \ .
\end{eqnarray*}
We rewrite this expression by reversing the order of summation: summing first over
$\vec{j}$\ then over $\vec{i}$.
\begin{eqnarray*}
\sum_{1\leq|\vec{i}|\leq N-1} \left.
\gamma^{\vec{i}}d_{\varepsilon}\tilde{C}_{\vec{i}}
\right|_{\tiny{P}} &=& \sum_{1<|\vec{j}| < N-1} \left. \left(
\sum_{0<\vec{i} \leq \vec{j}} \gamma^{\vec{i}} \frac{1}{(\vec{j} -
\vec{i})!} \frac{\partial^{|\vec{j} - \vec{i}|} C}{\partial
x_1^{j_1-i_1} \ldots \partial x_{M}^{j_M-i_M}} \right)
d_{\varepsilon} \varepsilon_{\vec{j}} \right|_{\tiny{P}}\\
&& + \sum_{N-1 \leq |\vec{j}| \leq N} \left. \left( \sum_{0<\vec{i}
< \vec{j}} \gamma^{\vec{i}} \frac{1}{(\vec{j} - \vec{i})!}
\frac{\partial^{|\vec{j} - \vec{i}|} C}{\partial x_1^{j_1-i_1}
\ldots \partial x_{M}^{j_M-i_M}} \right) d_{\varepsilon}
\varepsilon_{\vec{j}} \right|_{\tiny{P}} \ .
\end{eqnarray*}
We can now clearly read off the coefficient in front of each coordinate gradient $d_{\varepsilon} \varepsilon_{\vec{j}}$\ in the sum. Since each $d_{\varepsilon} \varepsilon_{\vec{j}}$\ is an
independent covariant coordinate vector, the coefficients must
vanish independently. For each $\vec{j}$\ satisfying $1<|\vec{j}| <
N-1$\ we then have a condition:
\begin{equation}\label{eq:coeffs1}
\sum_{ 0< \vec{i} \leq \vec{j}} \left. \gamma^{\vec{i}}
\frac{1}{(\vec{j} - \vec{i})!} \frac{\partial^{|\vec{j} - \vec{i}|}
C}{\partial x_1^{j_1-i_1} \ldots \partial x_{M}^{j_M-i_M}}
\right|_{\tiny{P}} = 0 \ ,
\end{equation}
while for $N-1\leq |\vec{j}| \leq N$\ we have
\begin{equation}\label{eq:coeffs2}
\sum_{ 0< \vec{i} < \vec{j}} \left. \gamma^{\vec{i}}
\frac{1}{(\vec{j} - \vec{i})!} \frac{\partial^{|\vec{j} - \vec{i}|}
C}{\partial x_1^{j_1-i_1} \ldots \partial x_{M}^{j_M-i_M}}
\right|_{\tiny{P}} = 0 \ .
\end{equation}
Now we assume that $P$\ lies on the classical constraint surface and
that the classical constraint is regular, i.e. $\left. C
\right|_{\tiny{P}} = 0$\ and $\left. dC \right|_{\tiny{P}} \neq 0$.
As an immediate consequence, terms proportional to $C$\ drop out in
the sum~(\ref{eq:coeffs1}) and thus one can use
condition~(\ref{eq:coeffs2}) for all $1<|\vec{j}|\leq N$. We proceed
in two steps: 1) show that~(\ref{eq:coeffs2}) implies that all
$\gamma^{\vec{i}}=0$\ for $|\vec{i}|=1$; 2) use induction to
conclude that
$\gamma^{\vec{i}}=0$\ for all $\vec{i}$. \\ \ \\

\emph{Step 1}. Since $\left. dC \right|_{\tiny{P}} \neq 0$, there is
a direction along which the derivative of $C$\ does not vanish at
$P$, i.e. for some $k\in \{1, 2, \ldots, M\}$, we have $\left.
\frac{\partial C}{\partial x_k} \right|_{\tiny{P}} \neq 0$. Denote
the coefficients $\gamma_{(k)}^m:= \gamma^{\vec{i}}$, where all but
the $k$-th entry in the $M$-tuple $\vec{i}$\ are zero, namely $i_n =
m \delta_{nk}$. For these coefficients, with $m \leq N$\ the
condition~(\ref{eq:coeffs2}) takes form
\begin{equation}\label{eq:condition2_spec}
\sum_{n=1}^{m-1} \left. \gamma_{(k)}^n \frac{1}{(m-n)!}
\frac{\partial^{m-n}C}{\partial x_k^{m-n}} \right|_{\tiny{P}} = 0 \ .
\end{equation}
From our assumptions and the above relation evaluated at $m=2$\ we
immediately conclude that $\gamma_{(k)}^1=0$. This is true for any
$k$\ such that $\left. \frac{\partial C}{\partial x_k}
\right|_{\tiny{P}} \neq 0$. Now consider any $l\neq k$. Denote the
coefficients $\gamma_{(k, l)}^{r\ s} = \gamma^{\vec{i}}$, where
$k$-th and $l$-th entries of the $M$-tuple $\vec{i}$\ are $r$\ and
$s$\ respectively, while the rest are zero, i.e. $i_n = r\delta_{kn}
+ s\delta_{ln}$. (Note that $\gamma_{(k, l)}^{r\ 0} =
\gamma_{(k)}^r$). Let $\vec{j}$\ have components $j_n = \delta_{kn}
+ \delta_{ln}$, the condition~(\ref{eq:coeffs2}) becomes
\[
\left. \gamma_{(k, l)}^{1\ 0} \frac{\partial C}{\partial x_l}
\right|_{\tiny{P}} + \left. \gamma_{(k, l)}^{0\ 1} \frac{\partial
C}{\partial x_k} \right|_{\tiny{P}} = 0 \ .
\]
The first term vanishes as $\gamma_{(k, l)}^{1\ 0} =
\gamma_{(k)}^1=0$, while $\left. \frac{\partial C}{\partial x_k}
\right|_{\tiny{P}} \neq 0$\ and it follows that $\gamma_{(k, l)}^{0\
1}=0$. Thus for all $\vec{i}$\ satisfying $|\vec{i}|=1$\
condition~(\ref{eq:coeffs2})
together with $C|_{\tiny{P}}=0$\ imply $\gamma^{\vec{i}}=0$.\\ \ \\

\emph{Step 2}. We now assume that $\gamma^{\vec{i}}=0$\ for all
$|\vec{i}| \leq m$. Evaluating the sum~(\ref{eq:coeffs2}) for some
$\vec{j}$\ such that $|\vec{j}|=m+2$, and dropping the terms
multiplied by $\gamma^{\vec{i}}$\ with $|\vec{i}| \leq m$\ we obtain
\begin{equation}\label{eq:n_plus_2}
\sum_{\tiny{\begin{array}{c} l=1\\ j_l \neq 0
\end{array}}}^M \left. \gamma^{\vec{j} - \vec{v}_{(l)}} \frac{\partial
C}{\partial x_l} \right|_{\tiny{P}} = 0 \ ,
\end{equation}
where $\vec{v}_{(l)}$\ is the $M$-tuple with a single non-zero unit
entry $\vec{v}_{(l)\ n} = \delta_{ln}$. Setting $\vec{j} =
(m+2)\vec{v}_{(k)}$, where $k$, as before, labels a direction along
which $\left. \frac{\partial C}{\partial x_k} \right|_{\tiny{P}}
\neq 0$, the sum~(\ref{eq:n_plus_2}) reduces to a single term
\[
\left. \gamma^{(m+1)\vec{v}_{(k)}} \frac{\partial C}{\partial x_k}
\right|_{\tiny{P}} = 0 \quad {\rm implying} \quad
\gamma^{(m+1)\vec{v}_{(k)}}=0 \ .
\]
Consider the following `step' that takes us from
$(m+1)\vec{v}_{(k)}$\ to another $M$-tuple of degree $(m+1)$:
subtract $\vec{v}_{(k)}$\ and add $\vec{v}_{(l)}$, for some $l\neq
k$. Naturally, the first step yields $m\vec{v}_{(k)} +
\vec{v}_{(l)}$\ and we note that any $M$-tuple of degree $(m+1)$ can
be reached by starting at $(m+1)\vec{v}_{(k)}$\ and taking up to
$(m+1)$\ such steps. Using $\gamma^{(m+1)\vec{v}_{(k)}}=0$, it is
straightforward to verify that $\gamma^{\vec{i}}=0$\ for all
$\vec{i}$\ that are one step away from $(m+1)\vec{v}_{(k)}$, we only
need to set $\vec{j} = (m+1)\vec{v}_{(k)} + \vec{v}_{(l)}$, and use~(\ref{eq:n_plus_2}) to get
\[
\left. \gamma^{m\vec{v}_{(k)}+\vec{v}_{(l)}} \frac{\partial C}{\partial x_k} \right|_{\tiny{P}} + \left. \gamma^{(m+1)\vec{v}_{(k)}} \frac{\partial C}{\partial x_l} \right|_{\tiny{P}} = 0 \ .
\]  
Since $\gamma^{(m+1)\vec{v}_{(k)}}=0$, this immediately yields $ \gamma^{m\vec{v}_{(k)}+\vec{v}_{(l)}} = 0$. The process can be continued iteratively: we notice that $\vec{i}$\
being $r$\ steps away from $(m+1)\vec{v}_{(k)}$\ implies $i_k =
(m+1-r)$. Suppose that $\gamma^{\vec{i}}=0$\ for all $\vec{i}$\ that
are up to $r$\ steps away from $(m+1)\vec{v}_{(k)}$. Pick some
$\vec{i}$\ that is exactly $r$\ steps away from
$(m+1)\vec{v}_{(k)}$, so that $i_k = (m+1-r)$, and
evaluate~(\ref{eq:n_plus_2}) with $\vec{j} = (\vec{i} +
\vec{v}_{(l)})$\ for any $l\neq k$
\[
\sum_{\tiny{\begin{array}{c} l'=1\\ i_{l'}+\delta_{ll'} \neq 0
\end{array}}} \left. \gamma^{\vec{i} + \vec{v}_{(l)} - \vec{v}_{(l')}} \frac{\partial
C}{\partial x_l} \right|_{\tiny{P}} = 0 \ .
\]
Since $l\neq k$, $(\vec{i} + \vec{v}_{(l)} - \vec{v}_{(l')})_k =
(m+1 - (r+\delta_{l'k}))$, so that all but one `$\gamma$' are exactly $r$\ steps away from $(m+1)\vec{v}_{(k)}$, and hence vanish, leaving us with 
\[
\left. \gamma^{\vec{i} - \vec{v}_{(k)}+ \vec{v}_{(l)}}
\frac{\partial C}{\partial x_k} \right|_{\tiny{P}} = 0 \quad {\rm
implying} \quad \gamma^{\vec{i} - \vec{v}_{(k)}+ \vec{v}_{(l)}}=0 \ . 
\]
From here it quickly follows that $\gamma^{\vec{i}}=0$\ for all
$\vec{i}$\ that are up to $(r+1)$\ steps away from
$(m+1)\vec{v}_{(k)}$. Thus by induction it first follows that
$\gamma^{\vec{i}}=0$\ for all $|\vec{i}|=m+1$\ and therefore, again inductively, for all
$|\vec{i}|\leq N$. This
completes the proof.\\ \ \\

\textbf{Corollary:} \emph{for the system of constraints truncated at
orders $N=2$\ or $N=3$, the set of gradients $\{
d_{\varepsilon}\tilde{C}_{\vec{i}} \}_{1\leq |\vec{i}|\leq N-1}$\ is
linearly independent everywhere, where $C$\ is regular.}\\ \ \\

This follows as at these orders, no terms proportional to $C$\
appear in the constraints $\{ C_{\vec{i}} \}_{1\leq |\vec{i}|\leq
N-1}$\ and linear dependence leads to~(\ref{eq:coeffs2})\ for all
$\vec{j}$\ hence the above proof applies regardless of the value
taken by $C$.\\ \ \\

While this result guarantees that there is some interesting range
of values for which the constraint functions are functionally
independent for sufficiently semiclassical states, it does not
provide a method for verifying whether the gradients are linearly
independent for a given set of values off the constraint surface
taken by $x_i$. Fortunately, the result can be made somewhat stronger.\\ \ \\

\textbf{Stronger statement}: \emph{for the system of constraints
truncated at order $N$ the set of gradients $\{
d_{\varepsilon}\tilde{C}_{\vec{i}} \}_{1\leq |\vec{i}|\leq N-1}$\ is
linearly dependent at some point $P$, with $C|_{\tiny{P}} \neq 0$,
only if $\left. \frac{\partial^{N-2} }{\partial
x_k^{N-2}}\left(\frac{1}{C} \right) \right|_{\tiny{P}} = 0$\ for
every $k$\ such that $\left. \frac{\partial C}{\partial x_k}
\right|_{\tiny{P}} \neq 0$.}\\ \ \\

\textbf{Proof}: we show that the above condition is necessary for
\emph{linear dependence} of the gradients $\{ d_{\varepsilon}\tilde{C}_{\vec{i}}
\}_{1\leq |\vec{i}|\leq N-1}$\ by demonstrating that its converse
implies \emph{linear independence}.\\ \ \\

Specifically, let $P$\ be such that $C|_{\tiny{P}}\neq 0$\ and
suppose there is some direction labeled $k$\ such that $\left.
\frac{\partial C}{\partial x_k} \right|_{\tiny{P}} \neq 0$\ as well
as $\left. \frac{\partial^{N-2} }{\partial
x_k^{N-2}}\left(\frac{1}{C} \right) \right|_{\tiny{P}} \neq 0$.
Suppose, as in the previous proof, that there are some coefficients
$\gamma^{\vec{i}}$\ such that
\[
\sum_{1\leq|\vec{i}|\leq N-1} \left. \gamma^{\vec{i}}
d_{\varepsilon} \tilde{C}_{\vec{i}} \right|_{\tiny{P}} = 0 \ .
\]
We repeat the steps from the previous proof leading up to the
relation~(\ref{eq:condition2_spec}). This time, since
$C|_{\tiny{P}} \neq 0$\ we need to consider both types of conditions~(\ref{eq:coeffs1}) and~(\ref{eq:coeffs2}). The former condition takes the form of a recursion relation for $m<N-1$
\begin{equation}\label{eq:condition1_spec}
\left. \gamma_{(k)}^m C \right|_{\tiny{P}} = - \sum_{n=1}^{m-1}
\left. \gamma_{(k)}^n \frac{1}{(m-n)!}
\frac{\partial^{m-n}C}{\partial x_k^{m-n}} \right|_{\tiny{P}} \ .
\end{equation}
For $m=N-1$\ and $m=N$\ relation~(\ref{eq:condition2_spec}) still holds, since terms proportional to $C$\ do not appear in~(\ref{eq:coeffs2}). Using $m=N-1$\ in~(\ref{eq:condition2_spec}) we obtain
\begin{equation}\label{eq:condition1_Nmin1}
\sum_{n=1}^{N-2} \left. \gamma_{(k)}^n \frac{1}{(N-n-1)!}
\frac{\partial^{N-n-1}C}{\partial x_k^{N-n-1}} \right|_{\tiny{P}} = 0 \ .
\end{equation}
As we prove in the appendix, (\ref{eq:condition1_spec}) implies that for $m<N-1$
\begin{equation} \label{eq:recursion_result}
\gamma_{(k)}^m = \left. \frac{\gamma_{(k)}^1 C}{(m-1)!}
\frac{\partial^{m-1}}{\partial x_k^{m-1}}\left( \frac{1}{C} \right)
\right|_{\tiny{P}} \ ,
\end{equation}
which, substituted into~(\ref{eq:condition1_Nmin1}), immediately gives
\[
0 = \left. \frac{\gamma_{(k)}^1}{(N-2)!}
\frac{\partial^{N-2}}{\partial x_k^{N-2}}\left( \frac{1}{C} \right)
\right|_{\tiny{P}} \ .
\]
Therefore, either $\gamma_{(k)}^1 =0$\ or $\left.\frac{\partial^{N-2}}{\partial
x_k^{N-2}} \left( \frac{1}{C} \right) \right|_{\tiny{P}} = 0$\ , and
by our assumption we conclude that $\gamma_{(k)}^1 =0$. Immediately, from~(\ref{eq:recursion_result}) we conclude that
$\gamma_{(k)}^m=0$\ for $1 \leq m <N-1$. Finally, we use the
relation~(\ref{eq:condition2_spec}) setting $m=N$\ to obtain
\[
\gamma_{(k)}^{N-1} \left. \frac{\partial C}{\partial x_k}
\right|_{\tiny{P}} = -\sum_{n=1}^{N-2} \left.\gamma_{(k)}^n
\frac{1}{(N-n)!} \frac{\partial^{N-n}C}{\partial x_k^{N-n}}
\right|_{\tiny{P}} \ .
\]
The right-hand side vanishes as all the terms are proportional to
$\gamma_{(k)}^{m}$\ with $1\leq m <N-1$, and hence
$\gamma_{(k)}^{N-1}=0$, which now accounts for all the coefficients
$\gamma_{(k)}^{m}$.

Let us now consider some $l \neq k$\ and specialize
condition~(\ref{eq:coeffs1}) to $\vec{j}$\ with $j_n = m\delta_{kn}
+ \delta_{ln}$. With the same notation as in the previous proof, we
obtain for $1<m+1<N-1$
\[
\sum_{n=1}^m \left. \gamma_{(k,l)}^{n\ 0} \frac{1}{(m-n)!}
\frac{\partial^{m-n+1} C}{\partial x_k^{m-n} \partial x_l}
\right|_{\tiny{P}} + \sum_{n=0}^m \left. \gamma_{(k,l)}^{n\ 1}
\frac{1}{(m-n)!} \frac{\partial^{m-n} C}{\partial x_k^{m-n}}
\right|_{\tiny{P}} =0 \ .
\]
The entire first sum vanishes since $\gamma_{(k,l)}^{n\ 0} =
\gamma_{(k)}^{n} = 0$\ as established earlier. The relation can then
be rewritten as a recursion relation for $\gamma_{(k,l)}^{m\ 1}$\
with $0<m<N-2$, which is essentially identical
to~(\ref{eq:condition1_spec})
\begin{equation}\label{eq:pairwise_recursion}
\left. \gamma_{(k,l)}^{m\ 1} C \right|_{\tiny{P}} = -
\sum_{n=0}^{m-1} \left. \gamma_{(k,l)}^{n\ 1} \frac{1}{(m-n)!}
\frac{\partial^{m-n} C}{\partial x_k^{m-n}} \right|_{\tiny{P}} \ .
\end{equation}
In complete analogy with the proof of the auxiliary result~(\ref{eq:aux2}) (the only difference being that the base step is now $m=1$), it follows that for $0<m<N-2$
\[
\gamma_{(k,l)}^{m\ 1} = \left. \frac{ \gamma_{(k,l)}^{0\ 1} C}{m!}
\frac{\partial^m }{\partial x_k^m} \left( \frac{1}{C} \right)
\right|_{\tiny{P}} \ ,
\]
when condition~(\ref{eq:coeffs2}) is evaluated for $\vec{j}$\ with
$j_n = m\delta_{kn} + \delta_{ln}$, this leads to
\[
 \left. \frac{ \gamma_{(k,l)}^{0\ 1} }{(N-2)!}
\frac{\partial^{N-2} }{\partial x_k^{N-2}} \left( \frac{1}{C}
\right) \right|_{\tiny{P}} = 0 \ .
\]
Once again, our assumption forces us to conclude that
$\gamma_{(k,l)}^{0\ 1}=0$\ for arbitrary $l$.\\ \ \\

We have now established that $\gamma^{\vec{i}} =0$\ for all
$|\vec{i}|=1$. The inductive extension of this result to
$|\vec{i}|<N-1$\ is straightforward with $C|_{\tiny{P}} \neq 0$.
Specifically, assume $\gamma^{\vec{i}} =0$\ for all $|\vec{i}| \leq
n$. We can rewrite~(\ref{eq:coeffs1}) for any $\vec{j}$\ such that
$|\vec{j}|=n+1$\ to obtain
\[
\left. \gamma^{\vec{j}} C \right|_{\tiny{P}} = -\sum_{0< \vec{i} <
\vec{j}} \left. \gamma^{\vec{i}} \frac{1}{(\vec{j} - \vec{i})!}
\frac{\partial^{|\vec{j} - \vec{i}|} C}{\partial x_1^{j_1-i_1}
\ldots \partial x_{M}^{j_M-i_M}} \right|_{\tiny{P}} \ .
\]
As the sum on the right is over $\vec{i}$\ of degree $n$\ and lower,
every term vanishes by the inductive assumption and
$\gamma^{\vec{i}}=0$\ follows for all $|\vec{i}|<N-1$. To eliminate
$\gamma^{\vec{i}}$\ with $|\vec{i}|=N-1$\ we are forced to use
condition~(\ref{eq:coeffs2}) instead and complete the inductive
proof in the same manner as in the previous proof, i.e. by first
proving $\gamma^{(N-1)\vec{v}_{(k)}}=0$\ then using steps to
eliminate other coefficients of degree $N-1$. This completes the
proof.\\ \ \\

A slight additional restriction on linear dependence immediately
follows.\\ \ \\

\textbf{Corollary}: \emph{there is no open neighborhood in
$\Gamma_Q$\ in which the set of (truncated) gradients $\{
d_{\varepsilon}\tilde{C}_{\vec{i}} \}_{1\leq |\vec{i}|\leq N-1}$\
is everywhere linearly dependent.}\\ \ \\

This follows since $\left. \frac{\partial C}{\partial x_k}
\right|_{\tiny{P}} \neq 0$\ extends to some open neighborhood of
$P$\ by continuity, while $\frac{\partial^{N-2} }{\partial
x_k^{N-2}}\left(\frac{1}{C} \right) = 0$\ at best defines a
hypersurface of codimension 1.\\ \ \\

At this stage, one may hope that the results may be strengthened
further, however, unless additional restrictions are imposed, it is
possible to manufacture simple low-order polynomial constraints that
allow the gradients $\{ d_{\varepsilon}\tilde{C}_{\vec{i}} \}_{1\leq
|\vec{i}|\leq N-1}$\ to be linearly dependent on some closed subsets
of the values taken by $x_i$. Below we demonstrate this by an
explicit example. Consider the algebra generated by a single element
$\hat{x}$\ with the following constraint
\[
\hat{C} = (\hat{x}-\hat{\mathbf{1}})(\hat{x}^2+\hat{\mathbf{1}}) \ .
\]
The classical constraint has a single real root $x=1$, which we will
treat as the classical constraint surface. The constraint is regular
for all real values of $x$\ as $\frac{d}{dx}C=2x^2+(x-1)^2$. Since
the algebra is abelian the constraints have no reordering terms and
$C_{\vec{i}} = \tilde{C}_{\vec{i}}$.

Suppose we truncate the system at order $N=4$. Then, by the results
established above, the gradients $\{
d_{\varepsilon}\tilde{C}_{\vec{i}} \}_{1\leq |\vec{i}|\leq 3}$\ are
guaranteed to be linearly independent for any real value of $x$\ so
long as $\frac{d^2}{dx^2} \left( \frac{1}{C} \right) \neq 0$. It is
not difficult to see that
\[
\frac{d^2}{dx^2} \left( \frac{1}{C} \right) = \frac{1}{C^3} \left(
4x^2(3-4x+3x^2) \right) \ ,
\]
which vanishes at $x=0$. Indeed, one can verify directly that at
this order
\[
\left. \left( d_{\varepsilon}C_1 + d_{\varepsilon}C_2 \right)
\right|_{x=0} =0 \ ,
\]
where $C_i = \langle \widehat{\Delta x}^i \hat{C} \rangle$. At the
point where, in addition to $x=0$, all the moments vanish as well we
have $d_Q C_i|_{x=0,\ \varepsilon_j=0} = d_{\varepsilon}C_i$\ for
$i>1$\ and we conclude
\[
\left. \left( d_QC_1 + d_QC_2 \right) \right|_{x=0,\
\varepsilon_j=0} =0 \ ,
\]
and the truncated set of constraints is truly functionally dependent
at this point.

In particular, in this example the failure of $\{
d_{\varepsilon}\tilde{C}_{\vec{i}} \}_{1\leq |\vec{i}|\leq 3}$\ to
be everywhere linearly independent is not remedied when we take into
account the gradients with respect to the expectation values, which
vanish when moments are set to zero, nor can it be fixed by
accounting for the reordering terms proportional to $\hbar$, which
in this example do not appear. Instead, the situation may be helped
in two ways:
\begin{itemize}
\item By noticing that $x=0$, where the constraint conditions become
degenerate, is likely ``too far'' from the classical constraint
surface at $x=1$, and so we would not expect the semiclassical
truncation to be accurate for the expectation values lying so far
from $C=0$\ surface,
\item By noticing that for a polynomial $C$, and for some value
$x_0$\ the condition
\[
\left. \frac{d^{N-2}}{dx^{N-2}} \left( \frac{1}{C} \right)
\right|_{x=x_0} = 0 \ ,
\]
cannot hold for all $N$\ above a given order, thus truncating at
successively higher orders one eventually arrives at the set of
truncated constraint conditions that are functionally independent at
$x_0$. In this example, for instance $\left.\frac{d^3}{dx^3} \left(
\frac{1}{C} \right)\right|_{x=x_0}=0$\, but going one order higher
$\left.\frac{d^4}{dx^4} \left( \frac{1}{C}
\right)\right|_{x=x_0}=-24 \neq 0$, which guarantees independence at
$x=0$\ for truncation at $N=6$. This, of course, has the caveat that
there may now be other points at which the new set of conditions is
not independent.\\ \ \\
\end{itemize}

Finally, we would like to consider whether order $\hbar$\
contributions to $d_{\varepsilon} C_{\vec{i}}$\ can spoil linear
independence. This can happen if
\[
\sum_{1\leq|\vec{i}|\leq N-1} \left. \gamma^{\vec{i}}
d_{\varepsilon} \tilde{C}_{\vec{i}} \right|_{\tiny{P}} = O(\hbar) \ ,
\]
for coefficients $\gamma^{\vec{i}}$\ that are themselves of
semiclassical order $0$. Taking these terms into account, on the
classical constraint surface, we write
relation~(\ref{eq:condition2_spec}) for $m=2$\ as
\[
\left. \gamma_{(k)}^1 \frac{\partial C}{\partial x_k}
\right|_{\tiny{P}} =O(\hbar) \ ,
\]
which implies $\left. \frac{\partial C}{\partial x_k}
\right|_{\tiny{P}} =O(\hbar)$. Similarly, off the constraint
surface, the recursion relation~(\ref{eq:condition1_spec}) holds to
order $\hbar$\ and hence
\[
\gamma_{(k)}^m = \left. \frac{\gamma_{(k)}^1 C}{(m-1)!}
\frac{\partial^{m-1}}{\partial x_k^{m-1}} \left( \frac{1}{C} \right)
\right|_{\tiny{P}} + O(\hbar) \ ,
\]
and we get the condition
\[
\left. \frac{\gamma_{(k)}^1 C}{(N-2)!} \frac{\partial^{m}}{\partial
x_k^{m}} \left( \frac{1}{C} \right) \right|_{\tiny{P}} = O(\hbar) \ .
\]
Thus, unless $\frac{ C}{(N-2)!} \frac{\partial^{m}}{\partial
x_k^{m}} \left( \frac{1}{C} \right)$\ is `almost' zero, the
gradients $\{d_{\varepsilon} C_{\vec{i}}\}_{1\leq|\vec{i}|\leq
N-1}$\ inherit linear independence from their symmetric counterparts $\{ d_{\varepsilon}\tilde{C}_{\vec{i}} \}_{1\leq
|\vec{i}|\leq N-1}$.

\section{Conclusion}

So what \emph{is} a semiclassical state on a Lie algebra according to our construction? In a nutshell, it is a linear functional on the corresponding universal enveloping algebra that assigns ``increasingly small'' (as powers of $\hbar$) values to increasingly high order generalized moments of the generating elements (defined by~(\ref{eq:moments})). This definition is consistent with the physical intuition that a semiclassical wavefunction is ``sharply peaked'' about some classical state.

What use do we envisage for these semiclassical states? Assuming that a given quantum system is in a semiclassical state, truncating its space of states at some finite order in $\hbar$\ can dramatically simplify explicit computations of its physical properties. What remains is a truncated ``quantum phase space'', which, as demonstrated by the detailed analysis of Section~\ref{sec:PBOrder}, is equipped with a well-defined truncated ``quantum Poisson bracket'', representing the Lie bracket of the original Lie algebra. This bracket is of direct relevance to the dynamical evolution of quantum mechanics (see~(\ref{eq:Qevolution}) and the preceding discussion)---the truncated set of quantum dynamical equations is typically much simpler to solve numerically than the full Schr\"odinger equation (for an example see~\cite{EffCosm}).

For the type of quantum systems considered here (see Section~\ref{sec:QLieAlg}) central elements of the universal enveloping algebra represent redundancy conditions that must be imposed in the form of constraints. Section~\ref{sec:Constraints} explicitly demonstrates the consistency of reducing the constraint system with the use of the semiclassical truncation in the case of a single generator of the center. A related question, not treated in the present work (but see~\cite{Casimir}), is the order-by-order characterization of the full set of inequalities that the positivity condition $\langle \hat{a} \hat{a}^* \rangle\geq 0$\ imposes on the generalized moments. In many situations of physical interest one needs to enforce these conditions, however, it remains to be demonstrated that, analogously to the constraints, the resulting set of inequalities is consistently reduced when the quantum system is truncated.

Overall, we have shown that the effective equations based on the semiclassical approximation previously developed for the canonical quantum systems in~\cite{EffAc} and~\cite{EffEq2}, can be applied to systems based on a finite-dimensional Lie algebra by adding certain constraint conditions. This generalization, already employed in~\cite{Casimir}, has direct applications to quantum mechanics, quantum cosmology and gauge fields.

\section*{Acknowledgements}

The author would like to thank Martin Bojowald for encouragement, numerous helpful discussions, and comments on the draft version of this manuscript.

\appendix
\section{Auxiliary results}

Here, we first prove by induction that for any derivative operator $\frac{\partial}{\partial x}$ and for any smooth function $C$,
\begin{equation}\label{eq:aux1}
\frac{\partial^m}{\partial x^m} \left( \frac{1}{C} \right) = -
\sum_{n=0}^{m-1} {m \choose n} \frac{1}{C} \frac{\partial^{m-n}
C}{\partial x^{m-n}} \frac{\partial^n }{\partial x^n} \left(
\frac{1}{C} \right) \ .
\end{equation}
It is straightforward to verify for $m=1$
\[
\frac{\partial}{\partial x} \left( \frac{1}{C} \right) =
-\frac{1}{C^2} \frac{\partial C}{\partial x} = - \sum_{n=0}^{0} {1
\choose n} \frac{1}{C} \frac{\partial^{1-n} C}{\partial x^{1-n}}
\frac{\partial^n }{\partial x^n} \left( \frac{1}{C} \right) \ .
\]
Assume this holds up to some $m$. Verify for $m+1$
\begin{eqnarray*}
\frac{\partial^{m+1}}{\partial x^{m+1}} \left( \frac{1}{C} \right)
&=& \frac{\partial}{ \partial x} \left( \frac{\partial^{m}}{\partial
x^{m}} \left( \frac{1}{C} \right) \right) = - \sum_{n=0}^{m-1} {m
\choose n} \frac{\partial}{\partial x} \left(\frac{1}{C}
\frac{\partial^{m-n} C}{\partial x^{m-n}} \frac{\partial^n
}{\partial x^n} \left( \frac{1}{C} \right) \right) \\ &=&
\frac{1}{C} \frac{\partial C}{\partial x} \sum_{n=0}^{m-1} {m
\choose n} \frac{1}{C} \frac{\partial^{m-n} C}{\partial x^{m-n}}
\frac{\partial^n }{\partial x^n} \left( \frac{1}{C} \right) \\ &&-
\sum_{n=0}^{m-1} {m \choose n} \frac{1}{C} \frac{\partial^{m-n+1}
C}{\partial x^{m-n+1}} \frac{\partial^n }{\partial x^n} \left(
\frac{1}{C} \right) \\ &&- \sum_{n=0}^{m-1} {m \choose n}
\frac{1}{C} \frac{\partial^{m-n} C}{\partial x^{m-n}}
\frac{\partial^{n+1} }{\partial x^{n+1}} \left( \frac{1}{C}
\right)\\ &=& -\frac{1}{C} \frac{\partial C}{\partial x}
\frac{\partial^{m}}{\partial x^{m}} \left( \frac{1}{C} \right) -
\sum_{n=1}^{m-1} \left( {m \choose n-1} + {m \choose n} \right)
\frac{1}{C} \frac{\partial^{m-n} C}{\partial x^{m-n}}
\frac{\partial^n }{\partial x^n} \left( \frac{1}{C} \right)\\ &&
-\frac{1}{C^2} \frac{\partial^{m+1} C}{\partial x^{m+1}} - m
\frac{1}{C} \frac{\partial C}{\partial x}
\frac{\partial^{m}}{\partial x^{m}} \left( \frac{1}{C} \right) \\
&=& - \sum_{n=0}^{m} {m+1 \choose n} \frac{1}{C}
\frac{\partial^{(m+1)-n} C}{\partial x^{(m+1)-n}} \frac{\partial^n
}{\partial x^n} \left( \frac{1}{C} \right) \ ,
\end{eqnarray*}
as required.

Assuming $C|_{\tiny{P}} \neq 0$, we use the above to prove that from
the recursion relation~(\ref{eq:condition1_spec}) it follows that for
$m<N-1$\
\begin{equation}\label{eq:aux2}
\gamma_{(k)}^m = \left. \frac{\gamma_{(k)}^1 C}{(m-1)!}
\frac{\partial^{m-1}}{\partial x_k^{m-1}} \left( \frac{1}{C} \right)
\right|_{\tiny{P}} \ .
\end{equation}
Setting $m=2$\ in~(\ref{eq:condition1_spec}) we verify the base step
\[
\gamma_{(k)}^2 C|_{\tiny{P}} = -\gamma_{(k)}^1 \left. \frac{\partial
C}{\partial x_k} \right|_{\tiny{P}} \ , \quad {\rm so\ that} \quad
\gamma_{(k)}^2 = \left. \gamma_{(k)}^1 C \frac{\partial }{\partial
x_k} \left( \frac{1}{C} \right) \right|_{\tiny{P}} \ .
\]
Assume this holds up to some $m<N-2$. So that for $m+1<N-1$ we start
from~(\ref{eq:condition1_spec}) and obtain
\begin{eqnarray*}
\gamma_{(k)}^{m+1} &=& -\frac{1}{C} \sum_{n=1}^{m} \left.
\gamma_{(k)}^n \frac{1}{(m+1-n)!} \frac{\partial^{m-n+1} C
}{\partial x_k^{m-n+1}} \right|_{\tiny{P}}
\\ &=&  -\frac{1}{C} \sum_{n=1}^{m} \left. \frac{\gamma_{(k)}^1 C}{(n-1)!}
\frac{\partial^{n-1}}{\partial x_k^{n-1}} \left( \frac{1}{C} \right)
\frac{1}{(m+1-n)!} \frac{\partial^{m-n+1} C }{\partial x_k^{m-n+1}}
\right|_{\tiny{P}}
\\ &=& -\frac{\gamma_{(k)}^1 C}{m!} \sum_{n=0}^{m-1} \left. {m
\choose n} \frac{1}{C} \frac{\partial^{m-n} C}{\partial x^{m-n}}
\frac{\partial^n }{\partial x^n} \left( \frac{1}{C} \right)
\right|_{\tiny{P}} = \left. \frac{\gamma_{(k)}^1 C}{m!}
\frac{\partial^{m}}{\partial x_k^{m}} \left( \frac{1}{C} \right)
\right|_{\tiny{P}} \ .
\end{eqnarray*}

Finally, using~(\ref{eq:aux2}) and evaluating
relation~(\ref{eq:condition2_spec}) for $m=N-1$\ yields
\begin{equation}\label{eq:aux3}
0 = \sum_{n=1}^{N-2} \left. \gamma_{(k)}^n \frac{1}{(N-1-n)!}
\frac{\partial^{N-1-n}C}{\partial x_k^{N-1-n}} \right|_{\tiny{P}} =
\left. \frac{\gamma_{(k)}^1 C}{(N-2)!} \frac{\partial^{m}}{\partial
x_k^{m}} \right|_{\tiny{P}} \ .
\end{equation}


\end{document}